\documentclass[a4paper, 10 pt, conference, twocolumn]{article}   
\usepackage{graphics} 
\usepackage{epsfig} 
\usepackage{mathptmx} 
\usepackage{times} 
\usepackage{amsmath} 
\usepackage{amssymb}

\usepackage{graphicx}
\usepackage{footmisc}
\usepackage{cancel}
\usepackage{subfigure}
\date{}
\usepackage{geometry}
\usepackage{cancel}
\newcommand{\norm}[1]{\left\lVert#1\right\rVert}
\DeclareRobustCommand{\abinom}{\genfrac{\langle}{\rangle}{0pt}{}}
 
\title{\LARGE \bf
Koopman Analytical Modeling of Position and Attitude Dynamics: a Case Study for Quadrotor Control}

\author{Simone Martini$^{1}$, Kimon P. Valavanis$^{1}$, Margareta Stefanovic$^{1}$ 
\thanks{$^{1}$ ECE Department, D. F. Ritchie School of Engineering and Computer Science, University of Denver, Denver, CO, USA,
        {\tt\small Email: firstname.lastname@du.edu.}}%
 }

\begin{document}

\maketitle
\thispagestyle{empty}
\pagestyle{empty}

\begin{abstract}
This research presents a novel, analytical, Koopman Operator based formulation for position and attitude dynamics which can be used to derive control strategies for underactuated systems. Compared to data driven Koopman based techniques, the analytical approach presented in this work is model based and allows for an exact linear representation of the original nonlinear position and attitude dynamics. In fact, the resulting infinite dimensional model, defined in the lifted state space, is linear in the autonomous component and state dependent in the control. A boundary study is carried on to define the range of validity of the finite truncation of the Koopman based model followed by a controllability and stabilizability analysis to show the feasibility of employing the derived model for control system design. Compared to existing literature formulation, the presented model results in a better approximation of the original dyanmics using a more compact truncation of the lifted state space. Moreover, the model is derived using the Koopman approach on the entirety of the dynamics and does not require the need of angular velocity dynamic compensation. A case study involving an underactuated quadrotor unmanned aerial vehicle (UAV) is provided to show that, for practical use, a truncated subset of the infinite dimensional model, embeds most of the original nonlinear dynamics and can be used to design linear control strategies in the lifted space which results in nonlinear controllers in the original state space. The main advantages of the presented approach reside in the effective use of linear control strategies for nonlinear plats and the solution of the underactuation problem employing a single control loop.
\end{abstract}

\section{Introduction}
Rigid body dynamics is an essential tool for the study of mechanical systems. In robotics and automation, position and attitude dynamics equations have set the foundation for the modeling and control of a wide variety of systems, leading to huge technological advancements. In this work, a novel Koopman based linear model of position and attitude dynamics is presented.\\ As in many physical models, the choice of a proper coordinates system is crucial in representing the evolution of the system and can greatly simplify or compromise the finding of a solution. Using Newton's second law of motion  allows to represent the rigid body position and attitude dynamics as a set of nonlinear differential equations. Compared to linear systems, nonlinear models, add greater complexity and challenges in control system design since there is still no general framework for nonlinear control theory. The main techniques that have been developed to overcome these problems consist in linearizing the dynamic around equilibrium points, designing model based approaches, and feedback linearization which exactly linearizes the dynamics through the control action. Although many problem have been solved thanks to the above techniques, they are by no means a general approach. Linearization looses validity when the system is far from the equilibrium points, model based approaches require a detailed knowledge of the plant, and Feedback linearization is not always possible for underactuated systems. On this note, Koopman operator theory has recently gained popularity among researchers since it allows to represent the nonlinear dynamics as a linear evolution of observables (possibly nonlinear function in the state space), giving promising results in generalizing the well known linear system analysis to nonlinear systems. In the foundational work of 1931 \cite{koopman1931hamiltonian}, Koopman proved that the evolution of the nonlinear dynamics of Hamiltonian systems occurs by a linear operator when it is represented in Hilbert function space. The main implication of the work is that a nonlinear dynamics can be embedded into a linear evolution of an infinite set of observables. The application of the theory remained limited until, in \cite{mezic2004comparison, mezic2005spectral}, the dimensionality problem was addressed by analyzing the spectral properties of the Koopman operator and achieving a finite representation through the Koopman operator eigenfunctions. This was partly achieved due to the advancement of dynamic mode decomposition (DMD) \cite{schmid2010dynamic} which allowed to approximate the Koopman operator directly from data and gave a data driven connotation the application of the theory. Since then, researchers has started to contribute and apply Koopman operator theory at an exponential rate \cite{manzoor2023vehicular}. In particular, this new framework has recently found wide adoption in aerospace \cite{Abraham2, mamakoukas2022robust,folkestad2020data,folkestad2020episodic, folkestad2022koopman, jin2020pontryagin,zheng2023optimal,manaa2024koopman}  and robotics \cite{sinha2022koopman,goyal2022impedance,zhu2022koopman,haggerty2020modeling,bruder2021koopman,bruder2020data,Bruder,Bruder2}, field. As anticipated, most of the ongoing research revolves around data driven extraction of the Koopman operator or Koopman eigenfunction fueled by the increasing interest in model free solution for modeling and control \cite{kaiser2020data,Bevanda,Mezic,mezic2022numerical}. Only a very limited amount of works have proposed model based analytical solution to derive a set of Koopman observables or Koopman eigenfunctions. Some limited studies exploit model based known Koopman eigenfunction such as the Hamiltonian function, to embed the nonlinear dynamics into a lower dimensional space \cite{kaiser2021data}. In \cite{martini2024koopman} the authors use a single scalar Koopman eigenfunction to design the attitude controller of a quadrotor.  In \cite{Chen}, a set of analytically derived nonlinear measurement function is proposed for modeling and control of the attitude dynamics providing numerical simulations and experimental results. In \cite{Zinage}, a similar methodology is applied to derive the position and attitude dynamics model of a quadrotor using Euler angles configuration. A more compact set of observables to represent the dynamics of the quadrotors has been proposed by the authors in \cite{martini2023koopman} and used to design test the first analytically derived Koopman based quadrotor controller. An hybrid approach is adopted in \cite{Zinage2}, where a novel set of observables for position and attitude dynamics is analytically derived using dual quaternion representation and feed to Extended DMD (EDMD) to compute the lifted linear state and input matrix to be used for control design. The above methods clearly loose the advantages of a model free approach, however, using a set of model based analytically derived observables has been shown to provide better approximation of the original dynamics than the model free counterparts \cite{Zinage,Zinage2}. Nevertheless, all of the above studies do not use a complete Koopman based approach in that they all exploit dynamic compensation to simplify the dynamics of the angular velocity. In this paper, we present a novel set of Koopman observables to represent the linear evolution of position and attitude dynamics. Differently from other literature works, the presented observables include a set of Koopman observables that embeds the angular velocity dynamics and do not require the use of dynamic compensation. Although  several position and attitude representation (cylindrical coordinates, spherical coordinates, orientation matrix, orientation quaternion) are widely studied and relevant in the literature, the presented observables adopts the Euclidean coordinates and Euler angles representation as starting point. The Koopman based position and attitude model is accompanied by controllability analysis and a study of the validity bounds. Furthermore, a case study of using the derived model to design a control strategy for a quadrotor UAV is carried on. Results show that the coupled nonlinear dynamics embedded in the evolution of the observables allows to design a control law that solves the underactuation problem and allows to perform position and attitude control using a single control loop.

This article is structured as follow. The adopted notation and relevant background information are addressed in Section \ref{sec:2}. The novel set of Koopman based observables for the position and attitude dynamics is presented in Section \ref{sec:3} while the full derivation is referred in the Appendix. Section \ref{sec:4} addresses the controllability and boundary analysis followed by the numerical validation in Section \ref{sec:5}. A case study is carried on in Section \ref{sec:6} which exploits the presented Koopman based model to design a single loop trajectory tracking controller for the quadrotor. Finally, conclusion and future works are addressed in Section \ref{sec:7}.

\section{Notation and Background Information}\label{sec:2}
\subsection{Notation}
Denote $0_n$ and $I_n$ as the $n \times n$ zero matrix and identity matrix, respectively. Fix the unit vector in the third dimension as $\mathbf{e_3}= [0,0,1]^{\top}$. Given the vectors $a, b \in \mathbb{R}^{n}$ and matrices $A\in\mathbb{R}^{n\times n},B\in\mathbb{R}^{m\times m}$, let the operator ${\mathcal{T}}$ refer to the block transpose operation such that $[a,b]^{\mathcal{T}} \triangleq [a^{\top},b^{\top}]^{\top}$, define the operation $diag$ so that $diag(a) = a^{\top}I_n$, and denote the matrix $C=blkdiag(A,B)\in\mathbb{R}^{(n+m)\times(n+m)}$ as the block diagonal matrix with $A$ and $B$ diagonal elements. Let $S(\cdot)\in \mathbb{R}^{3\times 3}$ denote the skew symmetric matrix which can be used interchangeably to represent the cross product since, given $a, b \in \mathbb{R}^{3}$, $S(a)b \triangleq a \times b$.

\subsection{Position and Attitude Dynamics}
The Newton-Euler attitude and position dynamics in $SO(3)$ are presented as \cite{Lee}
\begin{align}
\dot R &= RS(\nu) \;,\\
J\dot \nu &= M - S(\nu)J\nu \label{attitudeso3}\\ 
\dot p &= Rv \;,\\
\dot v &= \frac{1}{m}F - S(\nu)v - gR^{\top}\mathbf{e_3} \;,\label{positionso3}
\end{align}
Where $R\in \mathbb{R}^{3\times 3}$ is the rotation matrix which depends on the Euler angles configuration of choice $\eta \in \mathbb{R}^{3}$, $\nu \in \mathbb{R}^{3}$ is the the angular velocity, $J\in \mathbb{R}^{3\times 3}$ is the inertia tensor, $M \in \mathbb{R}^{3}$ is the external torque, $p \in \mathbb{R}^{3}$ is the position vector, $v \in \mathbb{R}^{3}$ is the rigid body linear velocity expressed in the body fixed reference frame, $m$ is the total mass of the rigid body, $F \in \mathbb{R}^{3}$ is the external force vector, and $g$ is the gravitational acceleration.

\subsection{Koopman Operator}
Considering the continuous time nonlinear dynamics on a smooth $n$-dimensional manifold $\mathcal{M}$, the evolution of the state vector $x \in \mathcal{M}\subset\mathbb{R}^n$ is represented by
\begin{equation}\label{eqn:generalnon}
\frac{d}{{dt}}x(t) = f(x) \;,
\end{equation}
with corresponding smooth Lipschitz continuous flow $F_t:\mathcal{M}\mapsto\mathcal{M}$ 
\begin{equation}
F_t(x({t_0})) = x({t_0}) + \int_{{t_0}}^{{t_0} + t} {f(x(\tau ))} d\tau \;.
\end{equation}
with initial time $t_0\geq0$. The Koopman operator semigroup $\mathcal{K}_{t\geq0}$, referred from here on as the Koopman operator, acting on scalar function $b(x):\mathcal{M}\mapsto\mathbb{C}$, called observable, is defined on the state space $\mathcal{M}$ as
\begin{equation}
    \mathcal{K}_tb=b\circ F_t
\end{equation}
The Koopman operator is therefore an infinite dimensional linear operator that maps the evolution of the state observables \cite{Bevanda}. Moreover, an observable is called a Koopman eigenfunction $\phi$, associated to an eigenvalue $\lambda\in\mathbb{C}$, if
\begin{equation}
    (\mathcal{K}_t \phi)(x)=\phi(F_t(x))=e^{\lambda t}\phi(x)
\end{equation}
which is the solution of the linear ordinary differential equation 
\begin{equation}
    \frac{d}{{dt}}\phi(F_t(x))=\lambda\phi((F_t(x)))
\end{equation}
Finally, considering the eigenvalue $\lambda$ and the vector of Koopman observables $\varkappa = \left[\varkappa_1,\cdots,\varkappa_{n,\lambda}\right]^{\top}$, so that
\begin{equation}
    \varkappa(F_t(x)) = e^{J_{\lambda}t} \varkappa(x)
\end{equation}
and
\begin{equation}\label{eq:geneig}
    \frac{d}{{dt}}\varkappa(F_t(x)) = {J_{\lambda}} \varkappa(F_t(x))
\end{equation}
with $J_{\lambda}$ being the Jordan block 
\begin{equation}
   J_{\lambda} = \left[\begin{array}{ccc} \lambda & 1 &  \\ & \ddots & 1 \\ &  & \lambda\end{array}\right]
\end{equation}
then, the $\mathrm{span}\left\{\varkappa_1,\cdots,\varkappa_{n,\lambda}\right\}$ is a Koopman invariant subspace and $\varkappa$ is the vector of Koopman generalized eigenfunction \cite{9022864} that can be used to embed the original nonlinear dynamics of \eqref{eqn:generalnon}.

\section{Koopman Attitude \& Position Dynamics}\label{sec:3}
In this section, a set of observables, with the properties of generalized Koopman eigenfunction are analytically derived to represent the attitude and position dynamic of a rigid body. For brevity, only the final form is presented here while the full derivation steps are presented in the Appendix.
\subsection{Koopman Based Attitude Dynamics}
Considering (\ref{attitudeso3}), define the following vector observables of the state $\nu$
\begin{align}
    \nu_0 &=\nu, \;\;\;\;\; \nu_0 \in \mathbb{R}^3\\
    \gamma_0 &= J\nu,  \;\;\; \gamma_0 \in \mathbb{R}^3
\end{align}
Note that $\nu_0 = \nu$ is the state itself but it is here used for convenience. Following the recursive derivation of the nonlinear terms of \eqref{attitudeso3}, as done in the Appendix, the resulting dynamical model can be represented in the following Koopman quasi-linear form
\begin{align}
\dot \nu_k &= \nu_{k+1} + J^{-1}H_{k}(\nu)M\label{eq:nudyn}\\
\dot \gamma_k &= J\dot \nu_k
\end{align}
with
\begin{align}
\nu_k &= J^{-1}\sum_{n=0}^{k} \binom{k}{n} S(\gamma_n)\nu_{k-n}, \;\;\;\; \nu_k, \gamma_k \in \mathbb{R}^3
\end{align}
where $\binom{k}{n}$ is the binomial coefficient
\begin{equation}
    \binom{k}{n} = \frac{k!}{n!(k-n)!}
\end{equation}
Note that the emerging pattern resemble the one of a binomial expansion and general Leibniz rule. Moreover, 
\begin{align}\label{eqn:H}
H_{k+1}(\nu) &= \sum_{n=0}^{k} \binom{k}{n} h_nH_{k-n}, \;\;\;\; H_k\in \mathbb{R}^{3\times 3}\\
H_{0} &= I_3\\
h_k(\nu) &= S(\gamma_k)J^{-1} - S(\nu_k), \;\;\;\; h_k\in \mathbb{R}^{3\times 3}\label{eqn:h}
\end{align}
Considering $x_{\nu} = (\nu_0,\{\nu_k\}^{N_{\nu}-1}_{k=1}) \in \mathbb{R}^{n_{\nu}}$, where $n_{\nu} = 3\cdot N_{\nu}$ is the resulting system dimension, leads to the following system dynamics
\begin{align}
&\dot x_{\nu} = A_{\nu}x_{\nu} + B_{\nu}(x_{\nu})M\\
&A_{\nu} = \left [
    \begin{array}{cccc}
0 & [I_3] & \cdots & 0\\
\vdots & \vdots & \ddots & 0\\
0 & 0 & 0 & [I_3] \\
0 & 0 & 0 & 0 \\
    \end{array}
    \right],  \; \;
B_{\nu}(x_{\nu}) = \left [
\begin{array}{clll}
J^{-1}H_0\\
J^{-1}H_1\\
\vdots\\
J^{-1}H_{N_{\nu}-1}
\end{array}
\right]\nonumber\\ \label{eqn:Koopform1}
\end{align}
with $A_{\nu} \in \mathbb{R}^{n_{\nu}\times n_{\nu}}$ and $B_{\nu}(\nu) \in \mathbb{R}^{n_{\nu}\times 3}$. The system can be easily transformed into Jordan form using the proper transformation matrix, leading to the following state vector 
\begin{align}
    x_{\nu,J} &= (x_{\nu,J_1},x_{\nu,J_2},x_{\nu,J_3}) \\
    x_{\nu,J_i} &= (\nu_0(i), \cdots, \nu_{N_{\nu-1}}(i)), \;\;\;\; i = 1,2,3
\end{align}
which evolution is given by
\begin{align}
&\dot x_{\nu,J} = A_{\nu,J}x_{\nu,J} + B_{\nu,J}(x_{\nu,J})M \label{attitude_ss}\\
&A_{\nu,J} = \left [
    \begin{array}{cccc}
[A_{\nu,J_{1}}] & 0 & 0\\
0 & [A_{\nu,J_{2}}] & 0\\
0 & 0 &[A_{\nu,J_{3}}]\\
    \end{array}
    \right],  \; \;\\
&B_{\nu,J}(x_{\nu,J}) = \left [
\begin{array}{clll}
B_{\nu,J_1}\\
B_{\nu,J_2}\\
B_{\nu,J_3}
\end{array}
\right]\nonumber\\ \label{eqn:KoopJor}
\end{align}
where 
\begin{align}
 A_{\nu,J_{i}} &= \left[\!\!
    \begin{array}{ccccc}
0 & 1 & \cdots & 0\\
\vdots & \ddots & \ddots & 0\\
0 & 0 & 0 & 1\\
0 & 0 & 0 & 0\\
    \end{array}
    \!\!\right],  \; \;
B_{\nu,J_i} = \left[\!\!
\begin{array}{clll}
(J^{-1}H_0)(i,:)\\
(J^{-1}H_1)(i,:)\\
\vdots\\
(J^{-1}H_{N_{\nu}-1})(i,:)
\end{array}\!\!
\right]\nonumber\\ \label{eqn:Koopform}
\end{align}
with $A_{\nu,J_{i}} \in \mathbb{R}^{N_{\nu}\times N_{\nu}}$ and $B_{\nu,J_i} \in \mathbb{R}^{N_{\nu}\times 3}$.\\
Hence, the resulting system consists in a constant Jordan form state matrix, $A_{\nu,J}$, composed of three Jordan blocks, each corresponding to a zero eigenvalue with multiplicity $N_{\nu}$, and a state dependent control matrix $B_{\nu,J}(x_{\nu,J})$. Given the obtained structure, note that for the unforced system with ${N_{\nu}\rightarrow\infty}$, \eqref{eqn:Koopform} is of the form of \eqref{eq:geneig} and $x_{\nu,J_i}$ is a vector of Koopman generalized egenfunctions associated with the zero eigenvalue. Finally, each generalized eigenfunctions can be related to the respective angular velocity component $\nu(i)$.

\subsection{Koopman Based Position Dynamics}\label{sec:Kpd}
Selecting the same observables proposed in \cite{martini2023koopman}
\begin{align}
    g_0 &=R^{\top}g\mathbf{e_3},  \;\;\;\; g_0 \in \mathbb{R}^3\\
    v_0 &=R^{\top}\dot p,  \;\,\;\;\;\;\;\; v_0 \in \mathbb{R}^3\\
    p_0 &=R^{\top} p,  \;\;\;\;\;\;\;\; p_0 \in \mathbb{R}^3
\end{align}
in the next subsections, the full set of observables for the Koopman based position dynamics is derived by following the recursive derivation of the nonlinear terms of \eqref{positionso3}, as done in the Appendix.
\subsubsection{Derivation of $g_k$}
The $k$-th vector of gravitational observables dynamic is given by
\begin{equation}
\dot g_k = g_{k+1} - G_{k}M
\end{equation}
with
\begin{align}
g_{k+1} &= \sum_{n=0}^{k} \binom{k}{n} S^{\top}(\nu_{n})g_{k-n}
\end{align}
and 
\begin{align}\label{eqn:G}
G_k &= (G^{*}_k+G^{**}_k)\\
G^{*}_{k+1} &= \sum_{n=0}^{k} \binom{k}{n} S^{\top}(g_n)J^{-1}H_{k-n}\\
G^{**}_{k+1} &= \sum_{n=0}^{k} \binom{k}{n} S^{\top}(\nu_n)J^{-1}G_{k-n}\\
G_{0} &= 0_3
\end{align}
with $g_k \in \mathbb{R}^3$ and  $G_k\in \mathbb{R}^{3\times 3}$.
The observables are grouped in $x_{g} = (g_0,\{g_k\}^{N_{g}-1}_{k=1}) \in \mathbb{R}^{n_{g}}$, where $n_{g} = 3\cdot N_{g}$.\\
Assuming the gravitational acceleration as constant and choosing an appropriate rotation matrix to represent the attitude in terms of Euler angles, $g_0$ effectively embeds 2 two out of three Euler angles and $g_k$ their respective dynamics. To show this, consider the rotation matrix
\begin{equation}
\begin{array}{*{20}{c}}
R=\!{\left[ {\begin{array}{*{20}{c}}
{c_{\theta}c_{\psi}}&{s_{\varphi}s_{\theta}c_{\psi} - c_{\varphi}s_{\psi}} & {c_{\varphi}s_{\theta}c_{\psi} + s_{\varphi}s_{\psi}}\\
{c_{\theta}s_{\psi}}&{s_{\varphi}s_{\theta}s_{\psi} + c_{\varphi}c_{\psi}} & {c_{\varphi}s_{\theta}s_{\psi} - s_{\varphi}c_{\psi}}\\
{ - s_{\theta}}&{s_{\varphi}c_{\theta}} & {c_{\varphi}c_{\theta}}
\end{array}} \right]}
\end{array}
\end{equation}
which leads to $g_0 = g[c_{\theta},-c_{\theta}s_{\varphi},-c_{\theta}c_{\varphi}]^{\top}$. Hence, knowing $\nu$ and $g_0$ it's enough to compute the Euler rates from the relationship
\begin{equation}
\begin{array}{l}
\nu = \underbrace {\left[ {\begin{array}{*{20}{c}}
1&0&{ - s_{\theta}}\\
0&{c_{\varphi}}&{c_{\theta}s_{\varphi}}\\
0&{ - s_{\varphi}}&{c_{\theta}c_{\varphi}}
\end{array}} \right]}_W\dot\eta
\end{array}
\end{equation}
\subsubsection{Derivation of $v_k$}
The $k$-th vector of velocity observables dynamic is given by
\begin{equation}
\dot v_k = v_{k+1} + g_k - V_{k}M +  \frac{1}{m}\Omega_k F
\end{equation}
with
\begin{align}
v_{k+1} &= \sum_{n=0}^{k} \binom{k}{n} S^{\top}(\nu_{n})v_{k-n}
\end{align}
and
\begin{align}\label{eqn:V}
V_k &= (V^{*}_k+V^{**}_k)\\
V^{*}_{k+1} &= \sum_{n=0}^{k} \binom{k}{n} S^{\top}(v_n)J^{-1}H_{k-n}\\
V^{**}_{k+1} &= \sum_{n=0}^{k} \binom{k}{n} S^{\top}(\nu_n)J^{-1}V_{k-n}\\
V_{0} &= 0_3\\
\Omega_{k+1} &= \sum_{n=0}^{k} \binom{k}{n} S^{\top}(v_n)\Omega_{k-n}
\end{align}
with $v_k \in \mathbb{R}^3$, $V_k\in \mathbb{R}^{3\times 3}$, and $\Omega_k\in \mathbb{R}^{3\times 3}$.
The observables are grouped in $x_{v}\! = \!(v_0,\{v_k\}^{N_{v}-1}_{k=1}) \in \mathbb{R}^{n_{v}}$, where $n_{v}\! =\! 3\cdot N_{v}$.
\subsubsection{Derivation of $p_k$}
The $k$-th vector of position observables dynamic is given by
\begin{equation}
\dot p_k = p_{k+1} + v_k - P_{k}M
\end{equation}
with
\begin{align}
p_{k+1} &= \sum_{n=0}^{k} \binom{k}{n} S^{\top}(\nu_{n})p_{k-n}
\end{align}
and
\begin{align}\label{eqn:P}
P_k &= (P^{*}_k+P^{**}_k)\\
P^{*}_{k+1} &= \sum_{n=0}^{k} \binom{k}{n} S^{\top}(p_n)J^{-1}H_{k-n}\\
P^{**}_{k+1} &= \sum_{n=0}^{k} \binom{k}{n} S^{\top}(\nu_n)J^{-1}P_{k-n}\\
P_{0} &= 0_3
\end{align}
with $p_k \in \mathbb{R}^3$ and $P_k\in \mathbb{R}^{3\times 3}$.
The observables are grouped in $x_{p} = (p_0,\{p_k\}^{N_{p}-1}_{k=1}) \in \mathbb{R}^{n_{p}}$, where $n_{p} = 3\cdot N_{p}$.
\subsubsection{Low Order Koopman Based Positon Dynamics}\label{subsec:LOKoop}
Differently from the formulation in \cite{martini2023koopman}, in which the lifted state vector is obtained as $x_{\mathrm{tot}}= \left[x_g,x_v,x_p\right]^{\top}$, here, a lower dimensional solution is presented. Selecting 
\begin{equation}
    z_0 = p_0, \;\;\;\; z_0 \in \mathbb{R}^3
\end{equation}
Having defined $g_k, v_k$ and $p_k$, by applying the recursive derivation to the nonlinear terms of \eqref{positionso3}, an alternative arrangement leads to
\begin{align}
    \dot z_0 &= \underbrace{p_1+v_0}_{z_1}\\
    \dot z_1 &= p_2 + v_1 -P_1M +v_1 +g_0 +\frac{1}{m}\Omega_0F =\\
    &=\underbrace{p_2 + 2v_1 + g_0}_{z_2} - \underbrace{P_1}_{Z_1}M + \underbrace{\frac{1}{m}\Omega_0}_{\Xi_1}F\\
    \dot z_2 &= \underbrace{p_3 + 3v_2 + 3g_1}_{z_3} - \underbrace{(P_2+2V_1)}_{Z_2}M + \underbrace{2\frac{1}{m}\Omega_1}_{\Xi_2}F
\end{align}
\begin{align}
    \dot z_3 &= \underbrace{p_4 + 4v_3 + 6g_2}_{z_4} - \underbrace{(P_3+3V_2+3G_1)}_{Z_3}M + \underbrace{3\frac{1}{m}\Omega_2}_{\Xi_3}F\\
    \dot z_4 &= \underbrace{p_5 + 5v_4 + 10g_3}_{z_5} - \underbrace{(P_4+4V_3+6G_2)}_{Z_4}M + \underbrace{4\frac{1}{m}\Omega_3}_{\Xi_4}F
\end{align}
Hence, the $k$-th vector dynamics of an alternative position observables is given by 
\begin{equation}
    \dot z_k = z_{k+1} - Z_kM + \Xi_{k}F
\end{equation}
where, for $k\geq 2$,
\begin{align}
    z_k &= p_k + \alpha_k v_{k-1} + \beta_k g_{k-2} \\
    Z_k &= P_k + \alpha_k V_{k-1} + \beta_k G_{k-2}\\
    \Xi_k &= \alpha_k\frac{1}{m}\Omega_{k-1} \\
    \alpha_k &= k\\
    \beta_k &= \alpha_{k-1} + \beta_{k-1}\\
    \alpha_0 &= \beta_0 = 0 \\
    Z_0 &= 0_3\\
    \Xi_0 &= 0_3
\end{align}
with $z_k \in \mathbb{R}^3$, $Z_k\in \mathbb{R}^{3\times 3}$, and $\Xi_k \in \mathbb{R}^{3\times 3}$.\\
Considering $x_{z} = (z_0,\{z_k\}^{N_{z}-1}_{k=1}) \in \mathbb{R}^{n_{z}}$, where $n_{z} = 3\cdot N_{z}$ is the resulting system dimension, leads to the following system dynamics
\begin{align}
&\dot x_{z} = A_{z}x_{z} + B_{z}(x_{z})\zeta\\
&A_{z} = \left[\!
    \begin{array}{cccc}
0 & [I_3] & \cdots & 0\\
\vdots & \vdots & \ddots & 0\\
0 & 0 & 0 & [I_3] \\
0 & 0 & 0 & 0 \\
    \end{array}
    \!\right],  
&B_{z}(x_{z}) = \left[\!
\begin{array}{cc}
\Xi_0 \!\!\!&\!\!\! -Z_0\\
\Xi_1 \!\!\!&\!\!\! -Z_1\\
\vdots \!\!\!&\!\!\! \vdots\\
\Xi_{N_{z}} \!\!\!&\!\!v -Z_{N_{z}}
\end{array}
\!\right]\nonumber\\ \label{eqn:KoopPosform1}
\end{align}
with $\zeta = \left[F,M\right]^{\mathcal{T}} \in \mathbb{R}^{6}$, $A_{z} \in \mathbb{R}^{n_{z}\times n_{z}}$, and $B_{z}(x_z) \in \mathbb{R}^{n_{z}\times 6}$. As for the attitude dynamics, the system can be easily transformed into Jordan form using the proper transformation matrix, leading to the state vector 
\begin{align}
    x_{z,J} &= (x_{z,J_1},x_{z,J_2},x_{z,J_3}) \\
    x_{z,J_i} &= (z_0(i), \cdots, z_{N_{z-1}}(i)), \;\;\;\; i = 1,2,3
\end{align}
which evolution is given by
\begin{align}
&\dot x_{z,J} = A_{z,J}x_{z,J} + B_{z,J}(x_{z,J})\zeta \label{position_ss}\\
&A_{z,J} = \left [
    \begin{array}{cccc}
[A_{z,J_{1}}] & 0 & 0\\
0 & [A_{z,J_{2}}] & 0\\
0 & 0 &[A_{z,J_{3}}]\\
    \end{array}
    \right],  \; \;\\
&B_{z,J}(x_{z,J}) = \left [
\begin{array}{c}
B_{z,J_{1}}\\
B_{z,J_{2}}\\
B_{z,J_{3}}
\end{array}
\right]= \left [
\begin{array}{cc}
\Xi_{J_1} & -Z_{J_1}\\
\Xi_{J_2} & -Z_{J_2}\\
\Xi_{J_3} & -Z_{J_3}
\end{array}
\right]\nonumber\\ \label{eqn:KoopPosJor2}
\end{align}
where
\begin{align}
&A_{z,J_{i}} \!=\! \left[\!
    \begin{array}{ccccc}
0 & 1 & \cdots & 0\\
\vdots & \ddots & \ddots & 0\\
0 & 0 & 0 & 1\\
0 & 0 & 0 & 0\\
    \end{array}
    \!\right], B_{z,J_{i}} \!=\! \left[\!
\begin{array}{cc}
\Xi_0(i,:) \!\!\!&\!\!\! -Z_0(i,:)\\
\Xi_1(i,:) \!\!\!&\!\!\! -Z_1(i,:)\\
\vdots \!\!\!&\!\!\! \vdots\\
\Xi_{N_{z}}(i,:) \!\!\!&\!\!\! -Z_{N_{z}(i,:)}
\end{array}
\!\right]\nonumber\\ \label{eqn:KoopPosform}
\end{align}
with $A_{z,J_{i}} \in \mathbb{R}^{N_{z}\times N_{z}}$ and $B_{z,J_i} \in \mathbb{R}^{N_{z}\times 6}$.\\
Hence, the resulting system consists in a constant Jordan form state matrix, $A_{z,J}$, composed of three Jordan blocks, each corresponding to a zero eigenvalue with multiplicity $N_{z}$, and a state dependent control matrix $B_{z,J}(x_{z,J})$. Given the obtained structure, note that for the unforced system with ${N_{z}\rightarrow\infty}$, \eqref{eqn:KoopPosform} is of the form of \eqref{eq:geneig} and $x_{z,J_i}$ is a vector of Koopman generalized egenfunctions associated with the zero eigenvalue. Moreover, note that the position dynamics is coupled with the attitude dynamics and, in the case of null angular velocity and zero attitude angles, (\ref{positionso3}) is linear. FInally, each generalized eigenfunctions can be related to the respective coupled position component $z_0(i)$.
\subsection{Koopman Based Position and Attitude Dynamics}
The systems \eqref{attitude_ss} and \eqref{position_ss} are here combined to obtain the Koopman based rigid body dynamics model. The resulting state vector is
\begin{align}
     x = (\nu_0,\{\nu_k\}^{N_{\nu}-1}_{k=1},z_0,\{z_k\}^{N_{z}-1}_{k=1})
\end{align}
which evolution is given by

\begin{equation}
\dot x = Ax + B(x)\zeta \label{complete_ss}
\end{equation}
with
\begin{equation}
A = \left[\!\!
    \begin{array}{ccc}
[A_{z,J}] & 0\\
0 & [A_{\nu,J}]
    \end{array}
    \!\!\right], \;B(x) = \left[\!\!
\begin{array}{c}
B_{z,J}\\
\begin{array}{cc}
0_{n_\nu \times 3} & B_{\nu,J}
\end{array}
\end{array}
\!\! \right]\label{eqn:KoopAttPosJor}
\end{equation}

\section{Controllability and Boundary Analysis}\label{sec:4}
\subsection{Controllability Analysis}
The following considerations assume the inertia component to be non null $J_x,J_y,J_z\neq0$.
Given the formulation in \eqref{eqn:KoopAttPosJor} the $n-$dimensional pair $A,B(x(t))$ is controllable at time $t_0$ if and only if there exist a finite $t_1>t_0$ such that the controllability gramian
\begin{align}
    W_c(t_0,t_1)=\int^{t_1}_{t_0}e^{A\tau}B(x(\tau))B^{\top}(x(\tau))e^{A^{\top}\tau}d\tau
\end{align}
is nonsingular. From \cite{chen1984linear}, this is equivalent to verify that 
\begin{align}
    B^{\top}(x(\tau))e^{A^{\top}\tau}\textbf{v}\equiv\textbf{0}  \;\;\;\; \mathrm{or} \;\;\;\; \textbf{v}^{\top}e^{A\tau}B(x(\tau))\equiv\textbf{0} \label{eqn:ctrbcondition}
\end{align}
for all $\tau \in [t_0,t_1]$. Expressing $e^{A\tau}B(x(\tau))$ as a linear combination of $\{B(x(\tau)),AB(x(\tau)),\cdots,A^{n-1}B(x(\tau))\}$, an equivalent result to (\ref{eqn:ctrbcondition}) is 
\begin{align}
    \textbf{v}^{\top} \underbrace{\left[B(x(\tau))\;AB(x(\tau))\;\cdots\;A^{n-1}B(x(\tau))\right]}_{\mathcal{C}(x(\tau))}= \textbf{0}
\end{align}
meaning that the state dependent controllability matrix $\mathcal{C}(x(\tau))$ is full row rank. The latter, further implies that the matrix 
\begin{align}
    \left[A-\lambda I \; B(x(\tau))\right]
\end{align}
has full row rank at every eigenvalue $\lambda$ of $A$. Finally, with analog proof of Theorem $6.8$ in \cite{chen1984linear}, the Jordan form state space is controllable iff, the row of $B(x(\tau))$ corresponding to the last row of each Jordan blocks are linearly independent for all $x(\tau)$ with $\tau$ in $[t_0,t]$. Therefore, the Koopman based position and attitude dynamics is controllable iff 
\begin{align}
\left [
\begin{array}{cc}
\Xi_{N_{z}-1} & -Z_{N_{z}-1}\\
0_3 & J^{-1}H_{N_{\nu}-1}
\end{array}
\right] \label{eqn:ContrCheck}
\end{align}
is full rank.
Given the structure of (\ref{eqn:H}), it is easily verified that for $\nu=0$, $H_{N_{\nu}-1}(0)=0_{3\times3}$, hence the controllability is lost in the equilibrium point. Moreover, since (\ref{eqn:h}) is, by composition, always full rank for $\nu\neq0$, the matrix (\ref{eqn:H}) is constructed as a sum of full rank matrices which does not guarantee the full rank of the solution, hence no general proof for the rank of $H_{N_{\nu}-1}$ is given but a controllability check can be easily performed once the truncated system is computed. Finally, note that if the rigid body present symmetry, the inertia tensor will result in having two or all equal element in the diagonal, leading to a zero row in $H_k$ or $H_k=0_{3\times3}$ for $k>0$, respectively. In this case, the model is reduced by eliminating the resulting zero rows. This is evident when the inertia matrix is the identity matrix, leading to $S(\nu)J\nu=0$ and effectively rendering the original dynamics linear and the controllability proof trivial.

\subsection{Attitude Dynamics Boundary Analysis}\label{subsec:bounds}
In the following, unless otherwise specified, we consider the matrix and vector $\ell_2$ norm $\norm{\cdot}_2$ by omitting the subscript.
A relaxed upper bound on $\nu_{k+1}$ and $\gamma_{k+1}$ can be computed as
\begin{align}
    \norm{\nu_{k+1}} &\leq \underbrace{\norm{J^{-1}}\norm{\gamma}\norm{\nu} \sum^{k}_{n=0}\abinom{k+1}{n} (\norm{J^{-1}}\norm{\gamma})^n\norm{\nu}^{k-n}}_{\overline{\norm{\nu_{k+1}}}} \label{eq:boundnu}\\
    \norm{\gamma_{k+1}} &\leq \underbrace{\norm{\gamma}\norm{\nu}\sum^{k}_{n=0}\abinom{k+1}{n} (\norm{J^{-1}}\norm{\gamma})^n\norm{\nu}^{k-n}}_{\overline{\norm{\gamma_{k+1}}}}\label{eq:boundgamma}\nonumber\\
\end{align}
where $\abinom{k}{n}$ represent the Eulerian numbers. Equations \eqref{eq:boundnu},\eqref{eq:boundgamma} hold for $k>0$ with
\begin{align}
    \norm{\nu_0} &= \norm{\nu}\\
    \norm{\gamma_0} &= \norm{\gamma}
\end{align}
Noting that $\overline{\norm{\nu_{k+1}}} = \norm{J^{-1}} \overline{\norm{\gamma_{k+1}}}$, the following bound on $h_{k+1}$ is found
\begin{align}
    \norm{h_{k+1}} &= \norm{S(\gamma_{k+1})J^{-1} - S(\nu_{k+1})} \\\nonumber &\leq \left({\norm{J^{-1}}\norm{\gamma_{k+1}} + \norm{\nu_{k+1}}}\right) \\\nonumber &\leq 2\overline{\norm{\nu_{k+1}}}
\end{align}
for $k\geq0$ with
\begin{align}
    \norm{h_0} \leq \left({\norm{J^{-1}}\norm{\gamma} + \norm{\nu}}\right)
\end{align}
which leads to 
\begin{align}
    \norm{H_k} \leq \sum^{k}_{n=0}\abinom{k+1}{n} (\norm{J^{-1}}\norm{\gamma})^n\norm{\nu}^{k-n}\label{eq:boundH}
\end{align}
for $k\geq0$. Which leads to the bound 
\begin{align}
    \norm{\dot{\nu_k}} \leq \norm{H_k}\norm{J^{-1}}\left(\norm{\gamma}\norm{\nu} + \norm{M}\right)
\end{align}
Considering the upper bounds $\norm{\nu}\leq\norm{J^{-1}}\norm{\gamma}$, the above bounds can be relaxed into 
\begin{align}
    \norm{\nu_{k+1}} &\leq \left(\norm{J^{-1}}\norm{\gamma}\right)^{k+2}(k+1)!\label{eq:boundnurlx}\\
    \norm{H_{k}} &\leq \left(\norm{J^{-1}}\norm{\gamma}\right)^{k}(k+1)!\\
        \norm{\dot{\nu_k}} &\leq \left(\norm{J^{-1}}\norm{\gamma}\right)^{k}\!\!(k+1)!\!\left(\!\left(\norm{J^{-1}}\!\norm{\gamma}\right)^2 \!\!+\!\norm{J^{-1}}\!\norm{M}\!\right)\nonumber\\
        &\leq \left(\norm{J^{-1}}\norm{\gamma}\right)^{k+1}(k+1)!\left(\left(\norm{J^{-1}}\norm{\gamma}\right) +\frac{\norm{M}}{\norm{\gamma}}\right)\label{eq:boundnudotrlx}
\end{align}
whose terms will grow larger once $k>\frac{1}{\norm{J^{-1}}\norm{\gamma}}$, with $\norm{\gamma}\neq0$. The obtain results cannot guarantee that the truncation to a finite lower dimensional model would embed most of the nonlinear dynamics since the higher order derivatives do not tend to become smaller as $k\rightarrow\infty$. To ensure that the truncated model would approximate the original nonlinear dynamics, a numerical model validation is carried on in Section \ref{sec:5}.\\
Moreover, considering the Jordan form of $A_{\nu,J_{i}}$ in (\ref{eqn:Koopform}) and $t_0$ as starting time, the resulting solution of the unforced dynamics is
\begin{align}
   x_{\nu,J_i}(t) &= x_{\nu,J_i}(t_0)e^{A_{\nu,J_{i}}t}\\
   \nu(t) &= \nu_0(t) =\nu_0(t_0) + \nu_1(t_0)t + \cdots + \nu_k(t_0)\frac{t^k}{k!}
\end{align}
Which shows that, although $\nu_k$ would eventually be larger than $\nu_{k-1}$, it's weight on the unforced dynamics solution it's progressively smaller. Now, considering the norm of the solution and substituting (\ref{eq:boundnurlx}) yields to
\begin{align}
    \norm{\nu(t)} \leq &\norm{J^{-1}}\norm{\gamma(t_0)} + \left(\norm{J^{-1}}\norm{\gamma(t_0)}\right)^2t +\nonumber\\ &+ \cdots + \left(\norm{J^{-1}}\norm{\gamma(t_0)}\right)^{k+1}t^k\nonumber\\
    \leq &\norm{J^{-1}}\norm{\gamma(t_0)}(1 + \norm{J^{-1}}\norm{\gamma(t_0)}t +\nonumber\\ &+ \cdots + \left(\norm{J^{-1}}\norm{\gamma(t_0)}t\right)^{k})
\end{align}
which highlights that higher order terms become dominant for $t>t_{lim} =\frac{1}{\norm{J^{-1}}\norm{\gamma(t_0)}}$. So, it is expected that for $t>t_{lim}$, the error due to the truncation will have a greater compound leading to exponential divergence of the Koopman based model. In Section \ref{subsec:valid} it is heuristically shown that the Koopman attitude dynamical model is able to give a practically useful approximation of the original dynamics for $t<5t_{lim}$.
Considering (\ref{eq:boundnudotrlx}), we can assume that for forced systems in which $\frac{\norm{M}}{\norm{\gamma}}\ll\norm{J^{-1}}\norm{\gamma}$, $t_{lim}$ would still represent a valid bound. Otherwise a new bound for the forced dynamic has to bve computed. The solution of the forced dynamics is given by
\begin{align}
    x_{\nu,J}(t) &= x_{\nu,J}(t_0)e^{A_{\nu,J_{i}}t} + \int_{t_0}^{t}e^{A_{\nu,J}(t-\tau)}B_{\nu,J}(x_{\nu,J})M(\tau)\\
    \nu(t)& =\nu_0(t_0) + \cdots + \nu_k(t_0)\frac{t^k}{k!} +\nonumber\\
    &+\!J^{-1}\!\!\!\underbrace{\int_{t_0}^{t}\!\!\!\left(\!\!H_0\! + \!H_1(t-\tau)\! +\! \cdots \!+\! H_k\frac{(t-\tau)^k}{k!}\right)\!\!M(\tau)d\tau}_{B}
\end{align}
Applying the change of variables $p = t-\tau$ a relaxed bound for $\norm{B}$ can be found as
\begin{align}
    \norm{B} &= \norm{\int_{t}^{t_0}\!-\!\!\left(H_0\! + \!H_1\frac{p}{1!}\! +\! \cdots \!+\! H_k\frac{p^k}{k!}\right)\!\!\norm{M(t\!\!-\!\!p)}dp}\nonumber\\
    &\leq \int_{t}^{t_0}\!\!\!\left(\!\!\norm{H_0}\! + \!\norm{H_1}\!\!\frac{\norm{p}}{1!}\! +\! \cdots \!+\! \norm{H_k}\!\!\frac{\norm{p}^k}{k!}\!\right)\!\!\norm{M(t\!\!-\!\!p)}dp\nonumber\\
    &\leq \left[\!\!\left(\norm{H_0}\! + \!\norm{H_1}\!\!\frac{\norm{p}}{1!}\! +\! \cdots \!+\! \norm{H_k}\!\!\frac{\norm{p}^k}{k!}\!\right)\!\!\norm{M_{\int}(t\!\!-\!\!p)}\right]^{t_0}_{t}\!\!\!\!+\nonumber\\
    & - \!\!\int_{t}^{t_0}\!\frac{d\left(\norm{H_0}\! +\! \cdots \!+\! \norm{H_k}\frac{\norm{p}^k}{k!}\right)}{dp}\!\!\norm{M_{\int}(t\!\!-\!\!p)}dp\nonumber\\
    &\!\!\leq \left[\left(1\! + \!2\cancel{!}\norm{J^{-1}}\norm{\gamma}\frac{\norm{p}}{\cancel{1!}}\! +\! \cdots\!\right.\right.\nonumber\\
    &\left.\left.+\! (k+1)\cancel{!}\norm{J^{-1}}^k\norm{\gamma}^k\frac{\norm{p}^k}{\cancel{k!}}\right)\!\!\norm{M_{\int}(t\!\!-\!\!p)}\right]^{t_0}_{t}+\nonumber\\
    & - \!\!\int_{t}^{t_0}\!\frac{d\left(\norm{H_0}\! +\! \cdots \!+\! \norm{H_k}\frac{\norm{p}^k}{k!}\right)}{dp}\!\!\norm{M_{\int}(t\!\!-\!\!p)}dp\nonumber\\
\end{align}
Even without fully solving the above integral, it is noticeable that the higher order terms become relevant in the dynamics when
\begin{align}
    (k+1)\norm{J^{-1}}^k\norm{\gamma}^k\norm{t}^kU(t_0)>1\nonumber\\
    \left(\norm{J^{-1}}\norm{\gamma}\norm{t}\sqrt[k]{(k+1)M_{\int}(t_0)}\right)^k>1\nonumber
\end{align}
where $M_{\int}(t_0)$ is the integral of $M(t)$ evaluated in $t_0$. 
Hence, when $\frac{\norm{M}}{\norm{\gamma}}\cancel{\ll}\norm{J^{-1}}\norm{\gamma}$ the following indicative bound can be used
\begin{equation}\label{eq:boundt}
    t_{lim,M} = \frac{1}{\norm{J^{-1}}\norm{\gamma(t)}\sqrt[k]{(k+1)M_{\int}(t_0)}}
\end{equation}
\subsection{Complete Dynamics Boundary Analysis}\label{subsec:fullbounds}
Performing the boundary analysis on $a_k = g_k, v_k, p_k$ shows that 
\begin{equation}
    \norm{a_{k+1}}\leq \norm{a_0}\norm{\nu}\sum^{k}_{n=0}\abinom{k+1}{n}\left(\norm{J^{-1}}\norm{\gamma}\right)^n\norm{\nu}^{k-n}
\end{equation}
hence, a good approximation on the Koopman based position dynamics depends on the bounds computed for the attitude dynamics.
\section{Koopman Attitude and Position Dynamics Validation}\label{sec:5}
\subsection{Attitude Dynamics Numerical Validation}\label{subsec:valid}
\begin{figure}[htb!]
\centering
\includegraphics[width=1\columnwidth]{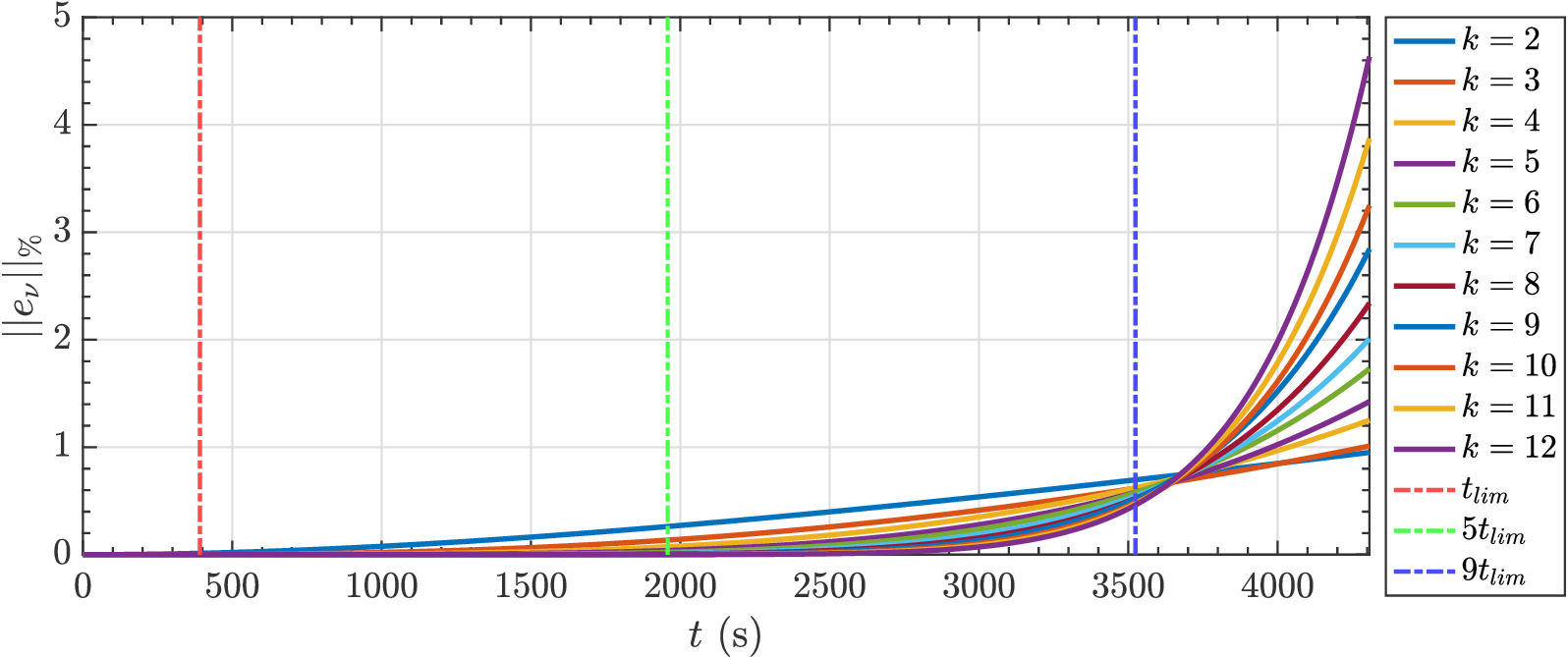}\hfill
\caption{\label{fig:1} Unforced approximation error with $J=J_0$ and $\nu(t_0) = 0.001$rad/s.}
\end{figure}

\begin{figure}[htb!]
\centering
\includegraphics[width=1\columnwidth]{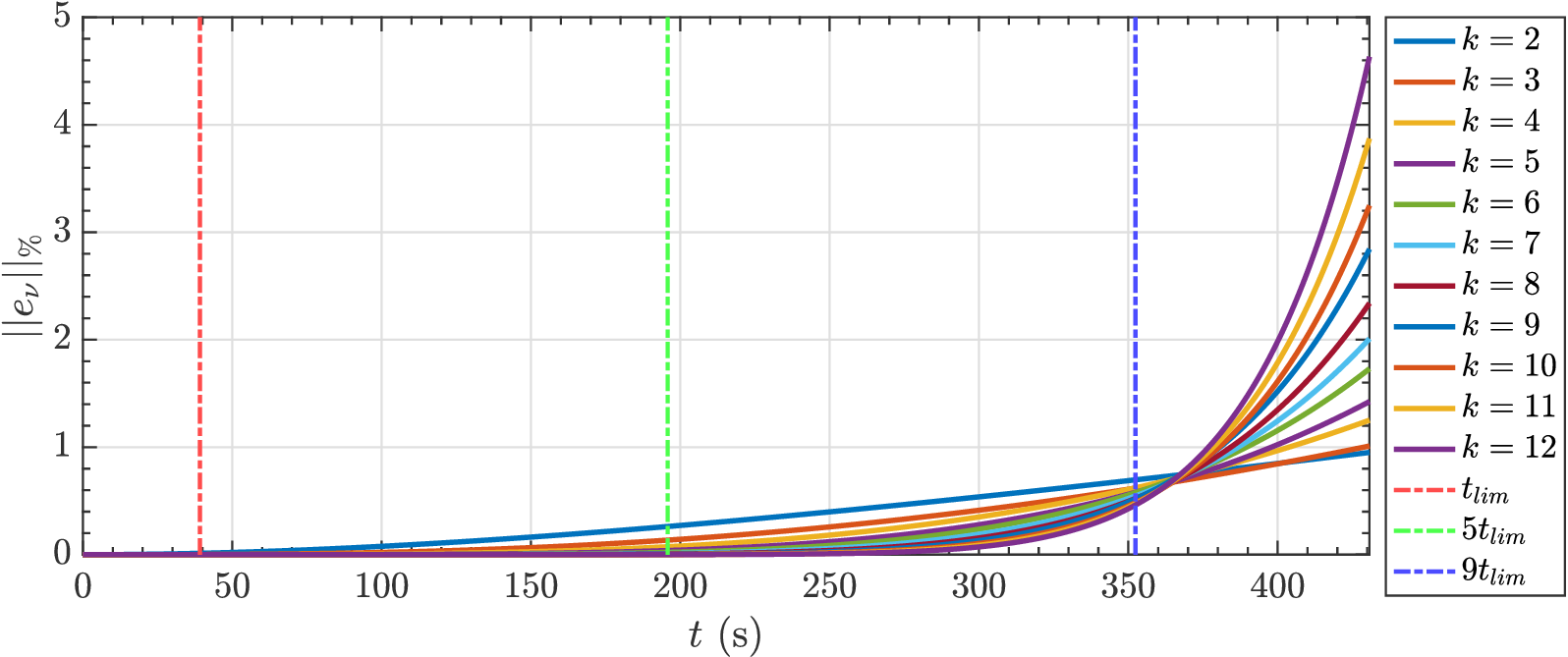}\hfill
\caption{\label{fig:2}  Unforced approximation error with $J=J_0$ and $\nu(t_0) = 0.01$rad/s.}
\end{figure}

\begin{figure}[htb!]
\centering
\includegraphics[width=1\columnwidth]{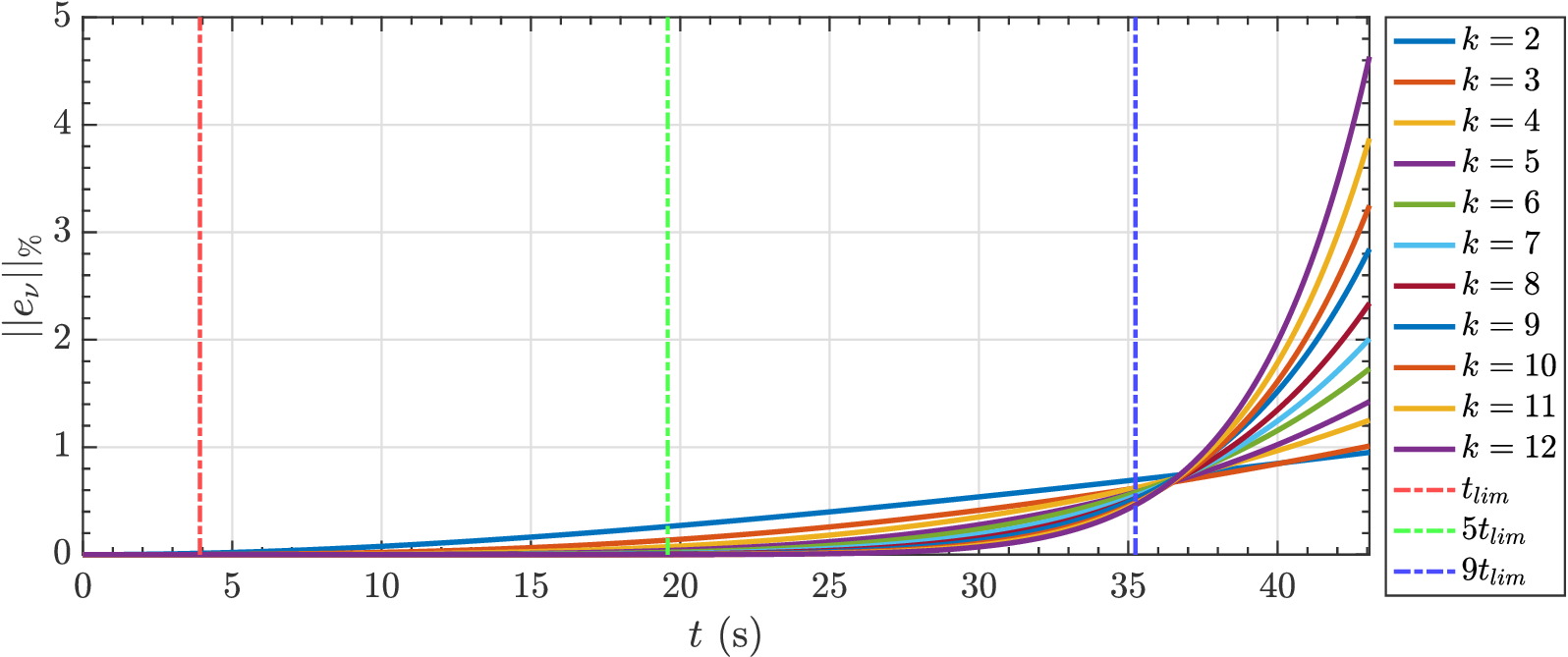}\hfill
\caption{\label{fig:3}  Unforced approximation error with $J=J_0$ and $\nu(t_0) = 0.1$rad/s.}
\end{figure}

\begin{figure}[htb!]
\centering
\includegraphics[width=1\columnwidth]{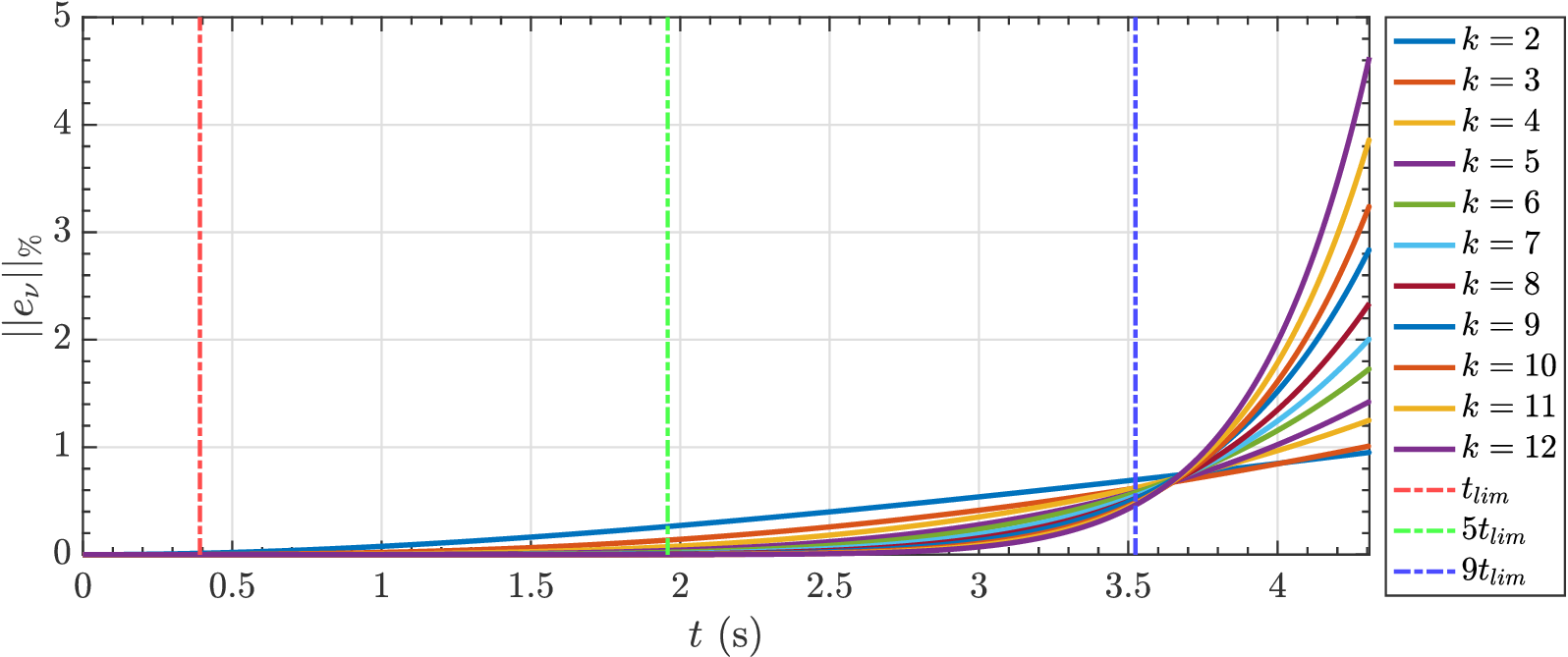}\hfill
\caption{\label{fig:4} Unforced approximation error with $J=J_0$ and $\nu(t_0) = 1$rad/s.}
\end{figure}

\begin{figure}[htb!]
\centering
\includegraphics[width=1\columnwidth]{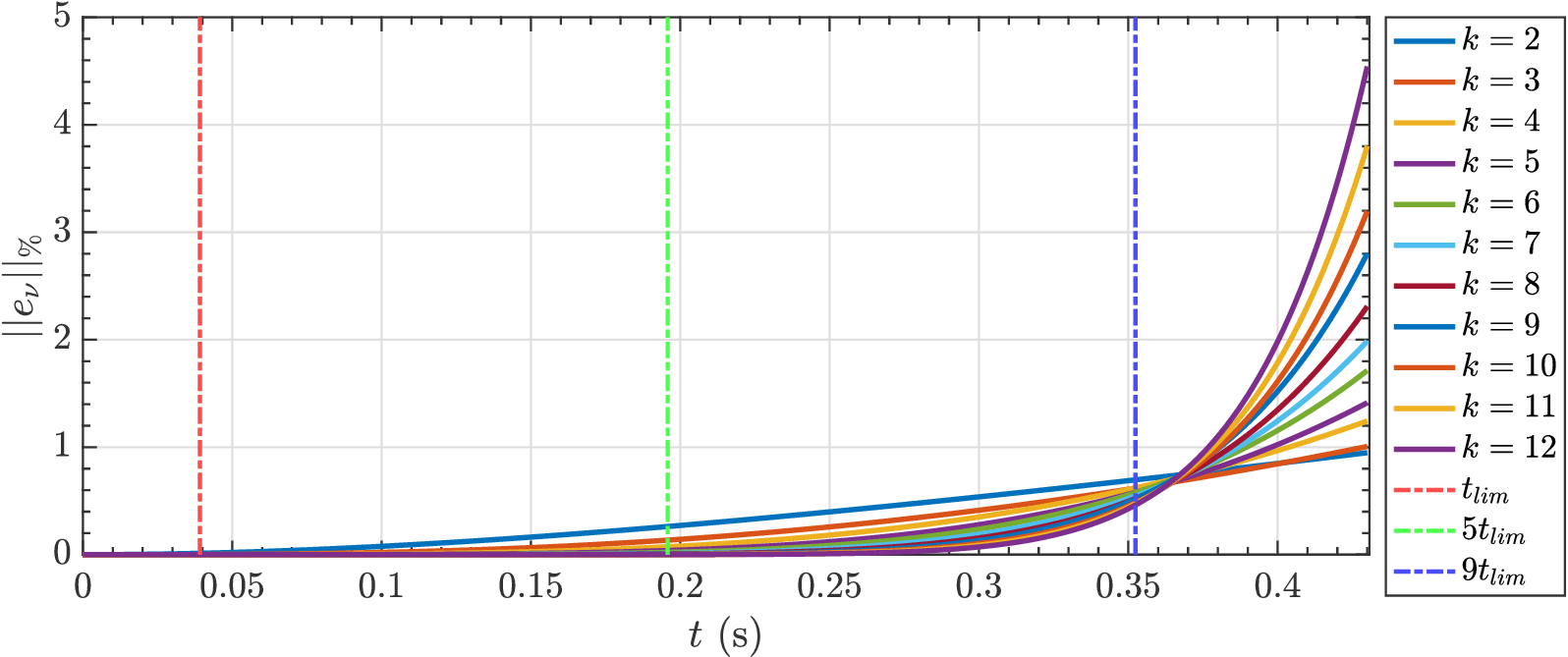}\hfill
\caption{\label{fig:5} Unforced approximation error with $J=J_0$ and $\nu(t_0) = 10$rad/s.}
\end{figure}

\begin{figure}[htb!]
\centering
\includegraphics[width=1\columnwidth]{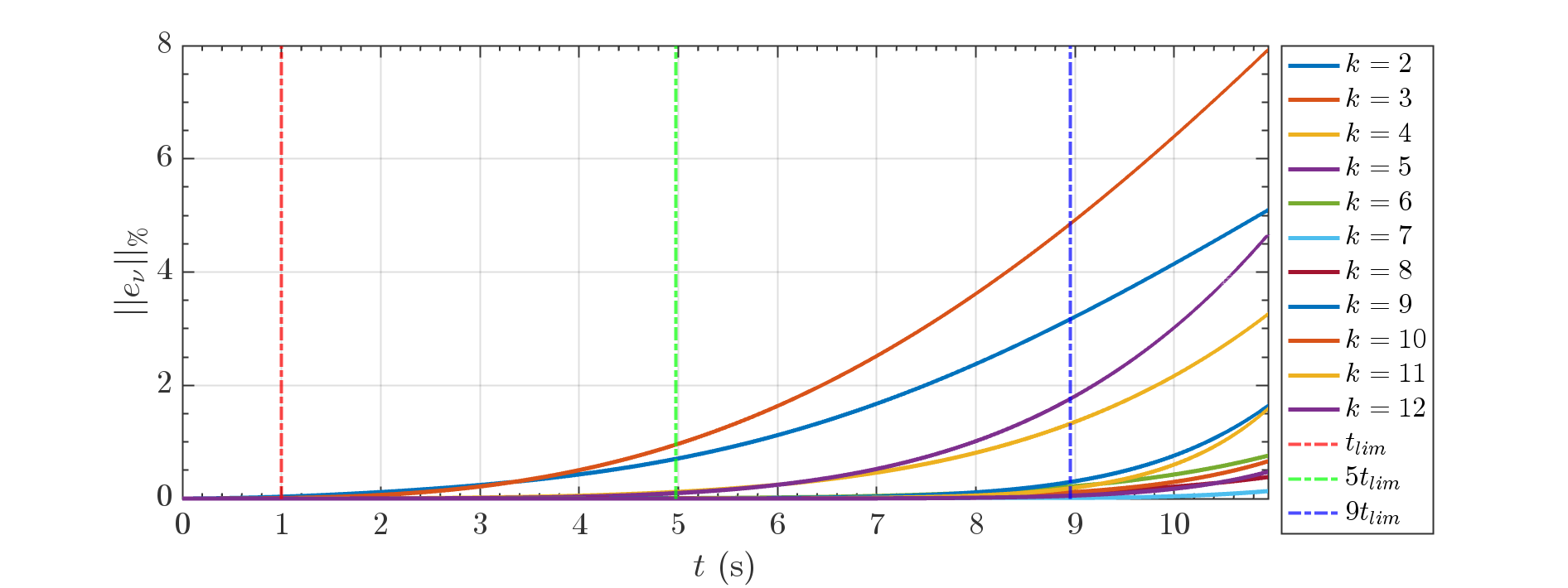}\hfill
\caption{\label{fig:6}  Unforced approximation error with $J=J_1$ and $\nu(t_0) = 0.001$rad/s.}
\end{figure}

\begin{figure}[htb!]
\centering
\includegraphics[width=1\columnwidth]{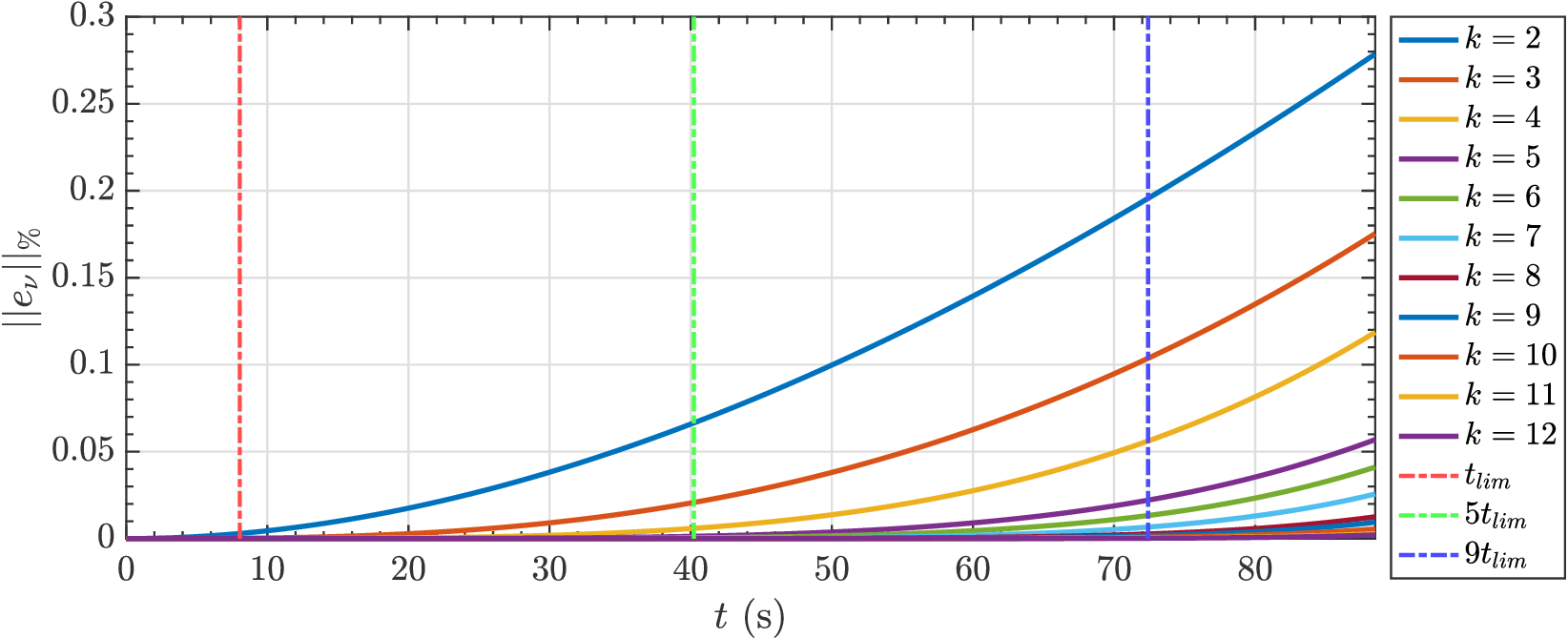}\hfill
\caption{\label{fig:7} Unforced approximation error with $J=J_2$ and $\nu(t_0) = 0.001$rad/s.}
\end{figure}

\begin{figure}[htb!]
\centering
\includegraphics[width=1\columnwidth]{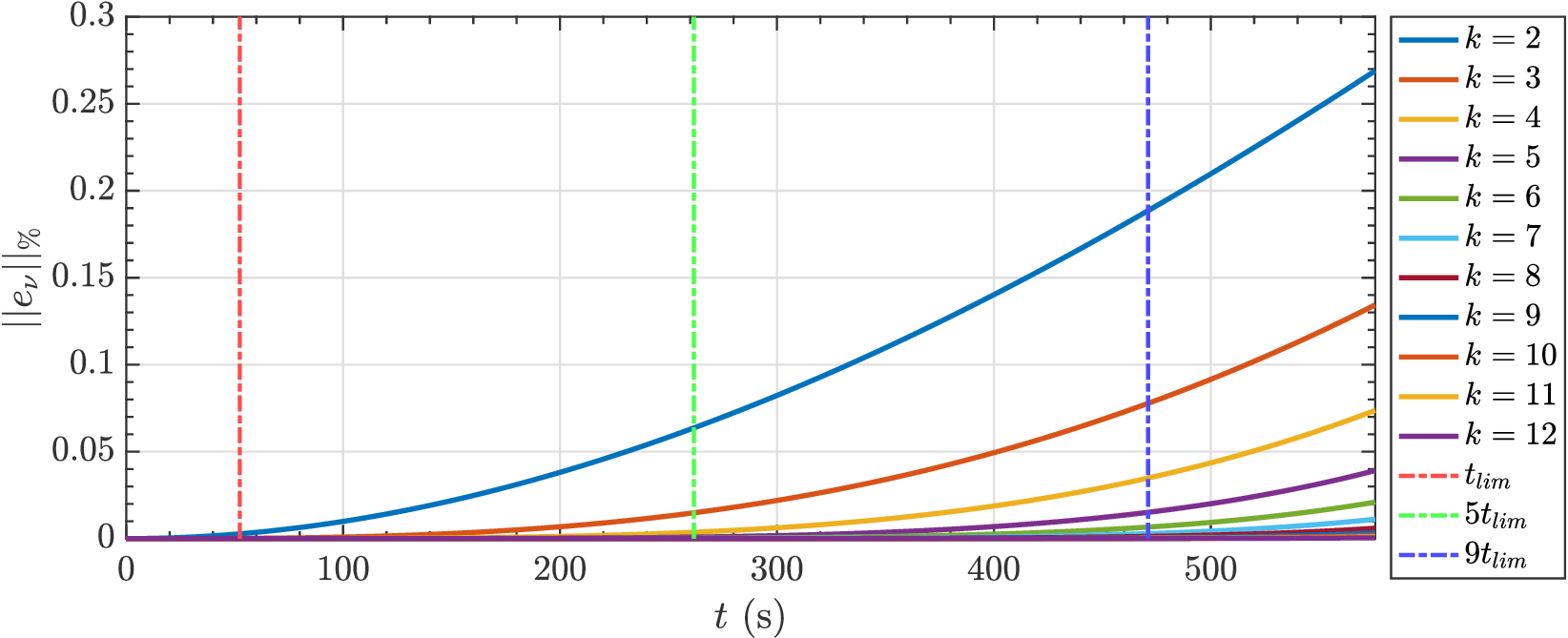}\hfill
\caption{\label{fig:8}  Unforced approximation error with $J=J_3$ and $\nu(t_0) = 0.001$rad/s.}
\end{figure}

\begin{figure}[htb!]
\centering
\includegraphics[width=1\columnwidth]{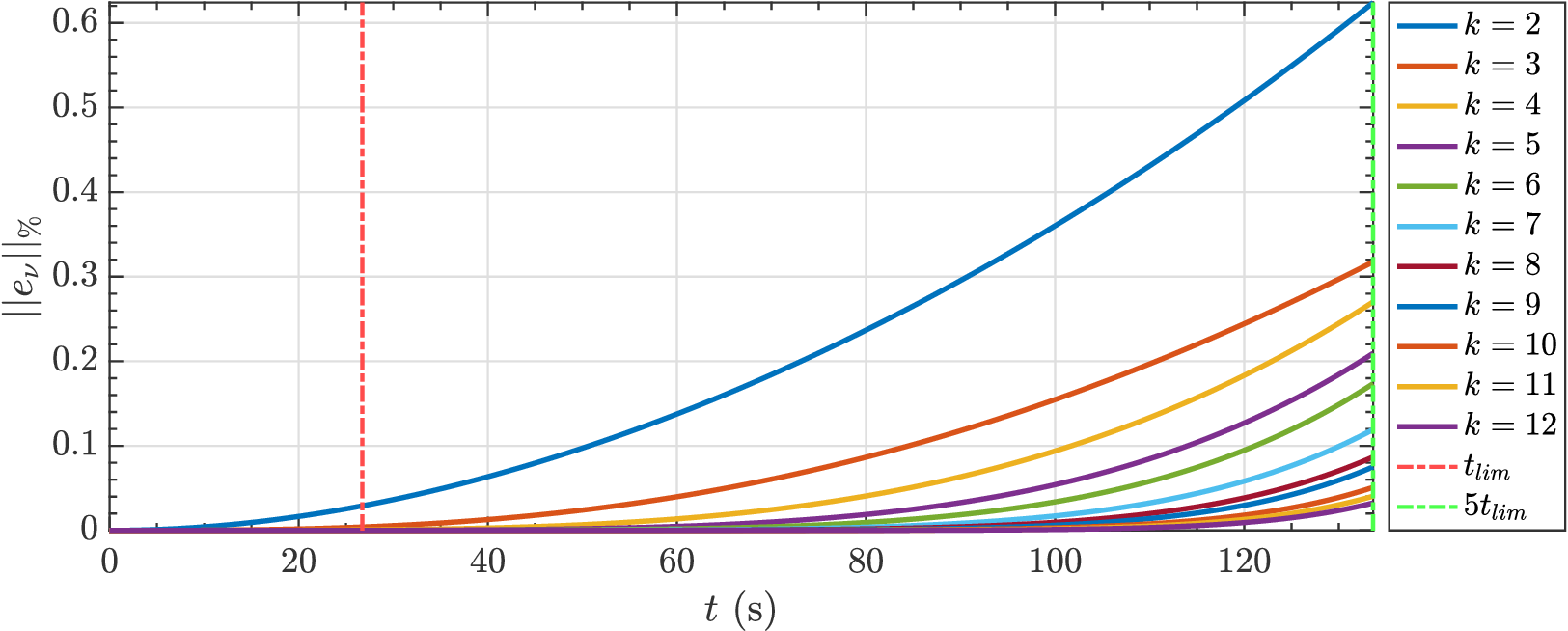}\hfill
\caption{\label{fig:9}  Unforced approximation error with $J=J_4$ and $\nu(t_0) = 0.001$rad/s.}
\end{figure}

In this section, the Koopman attitude model is compared to the original nonlinear model, first, by observing the evolution of the unforced dynamics, then, by the analysis of the open loop input response. The inertia matrix is set to $J = J_0 = diag(0.0131, 0.020, 0.0234)$ and the initial condition are progressively increased to reduce the value of $t_{lim}$. In Fig. \ref{fig:1}, \ref{fig:2}, \ref{fig:3}, \ref{fig:4}, and \ref{fig:5}, the approximation error is displayed for different number of observables. In this case the approximation error is computed by dividing the norm of the absolute error, between the nonlinear dynamics and the Koopman based model, by the norm of the initial angular velocity value
\begin{align}
    \norm{e_{\nu}}_{\%} = \frac{\norm{\nu_{nonlinear}-\nu_{Koopman}}}{\norm{\nu(t_0)}}
\end{align}
where $\nu_{nonlinear}$ and $\nu_{Koopman}$ are the angular velocity values computed by numerical integration of the nonlinear dynamics and the Koopman model, respectively.
In each of the figure, we can identify a time, $t_{switch}\approx 9*t_{lim}$, after which using a higher number of observables causes a larger approximation error. However before this time, increasing the number of observables drastically decrease the approximation error.
As expected from \eqref{eq:boundt}, using progressively larger initial angular velocity values, it is evident that $t_{switch}$ is inversely proportional to the initial angular velocity value.
To show the effect of the inertia tensor on $t_{lim}$, Fig. \ref{fig:6}, \ref{fig:7}, \ref{fig:8}, \ref{fig:9} present the $\norm{e_{\nu}}_{\%}$ of increasing number of observables, with the initial angular velocity value is set to $\nu(t_0)=0.001$ rad/s. The different tested inertia values are reported in Table \ref{tab:Inertia} and are chosen to represent distinct degrees of symmetry of the rigid body which can be related to the value of $\norm{J^{-1}}\norm{J}$. The results clearly show an increase of $t_{lim}$ along with an increase in symmetry of the rigid body. This is to be expected, since when the inertia has spherical symmetry, the nonlinear term in \eqref{attitudeso3} vanishes and the original dynamics becomes linear rendering it equivalent to the first observable of the Koopman based model. As further note, the value of $J_4$ in Fig. \ref{fig:9} is tested to show the validity of the model for $\norm{J}>1$.
\begin{table}[htb]
\caption{Inertia Values}
\begin{center}
\begin{tabular}{ccc}
\hline
\textbf{\textit{Param.}} & \textbf{\textit{Value}}  & \textbf{\textit{$\norm{J^{-1}}\norm{J}$}}                                                                                  \\ \hline
$J_1$         & $diag(0.001, 0.01, 0.1)$ Kgm$^2$     &    100          \\
$J_2$         & $diag(0.1, 0.11, 0.012)$ Kgm$^2$      &      9.1667           \\
$J_3$      & $diag(0.1, 0.11, 0.12)$ Kgm$^2$     & 1.2   \\    
$J_4$     & $diag(1, 2, 3)$ Kgm$^2$ & 3 \\\hline
\end{tabular}
\label{tab:Inertia}
\end{center}
\end{table}
Considering all the performed testing and additional ones with randomized inertia tensor and initial conditions, in the case of unforced dynamics, a time limit of $t<5t_{lim}$ allowed for a good approximation of the original nonlinear dynamics which increases with the number of observables.

\begin{figure}[htb!]
\centering
\includegraphics[width=1\columnwidth]{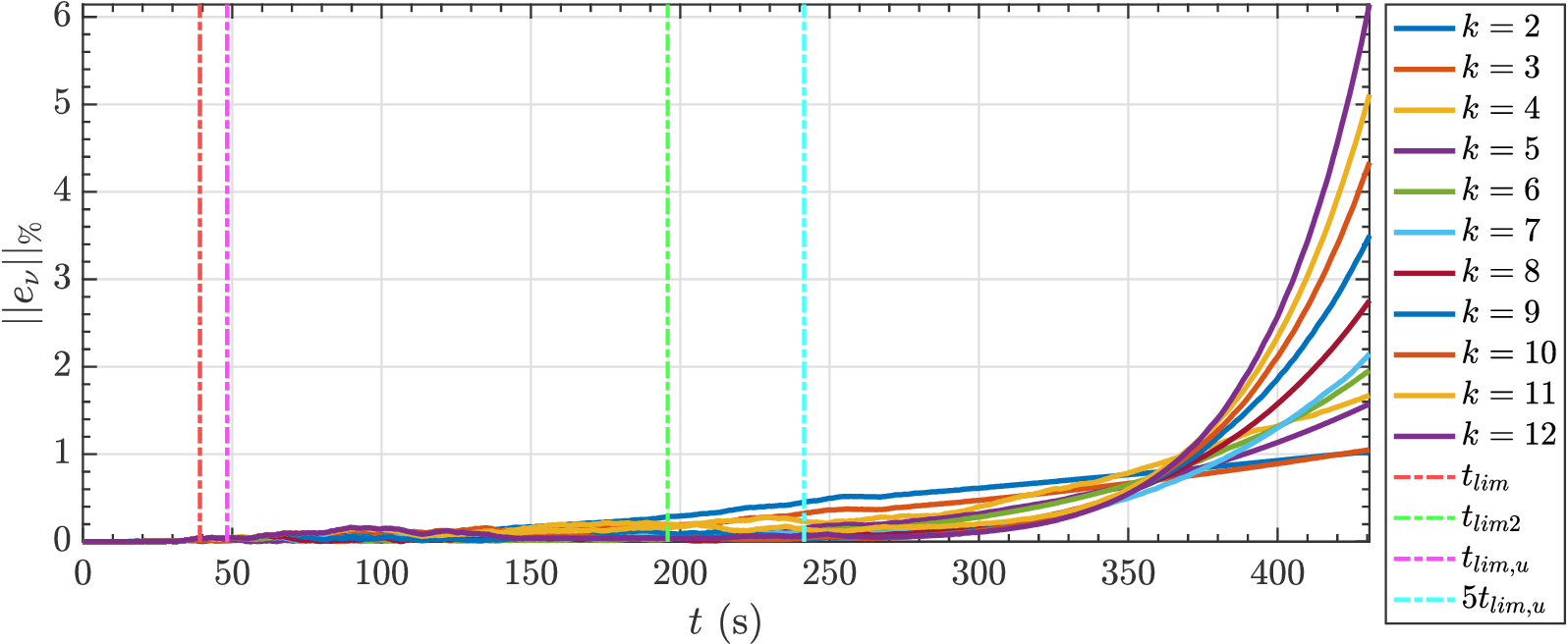}\hfill
\caption{\label{fig:10}  Forced approximation error with $J=J_0$, $\nu(t_0) = 0.01$rad/s, and $\alpha = 10^{-6}$.}
\end{figure}

\begin{figure}[htb!]
\centering
\includegraphics[width=1\columnwidth]{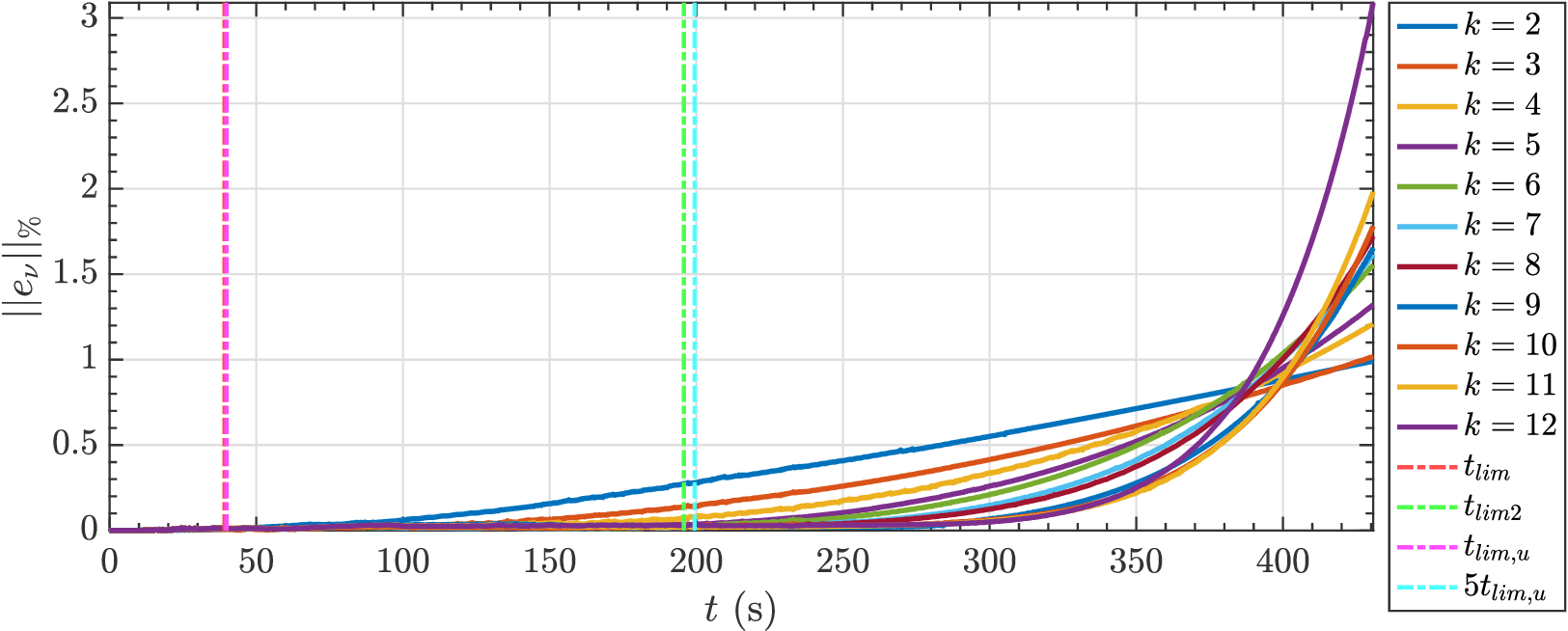}\hfill
\caption{\label{fig:11} Forced approximation error with $J=J_0$, $\nu(t_0) = 0.01$rad/s, and $\alpha = 10^{-5}$.}
\end{figure}

\begin{figure}[htb!]
\centering
\includegraphics[width=1\columnwidth]{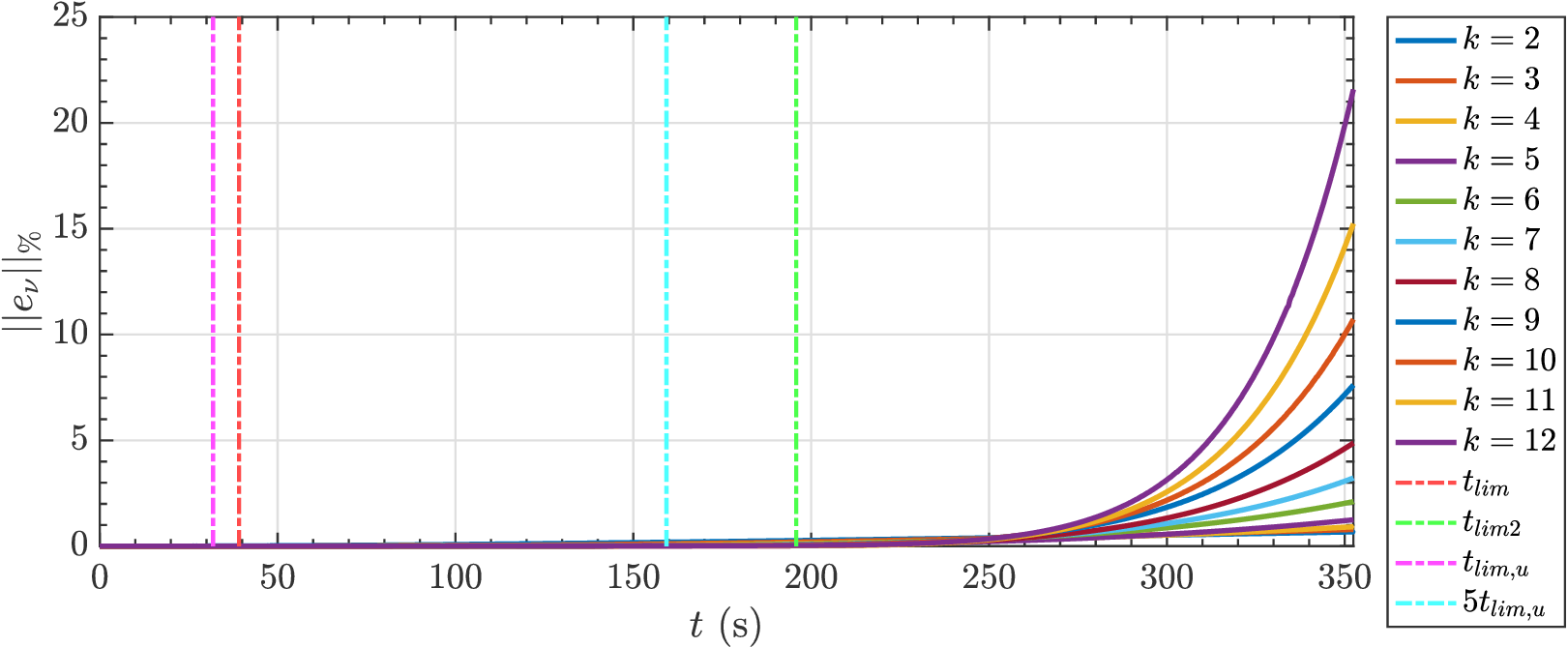}\hfill
\caption{\label{fig:12} Forced approximation error with $J=J_0$, $\nu(t_0) = 0.01$rad/s, and $\alpha = 10^{-4}$.}
\end{figure}

\begin{figure}[htb!]
\centering
\includegraphics[width=1\columnwidth]{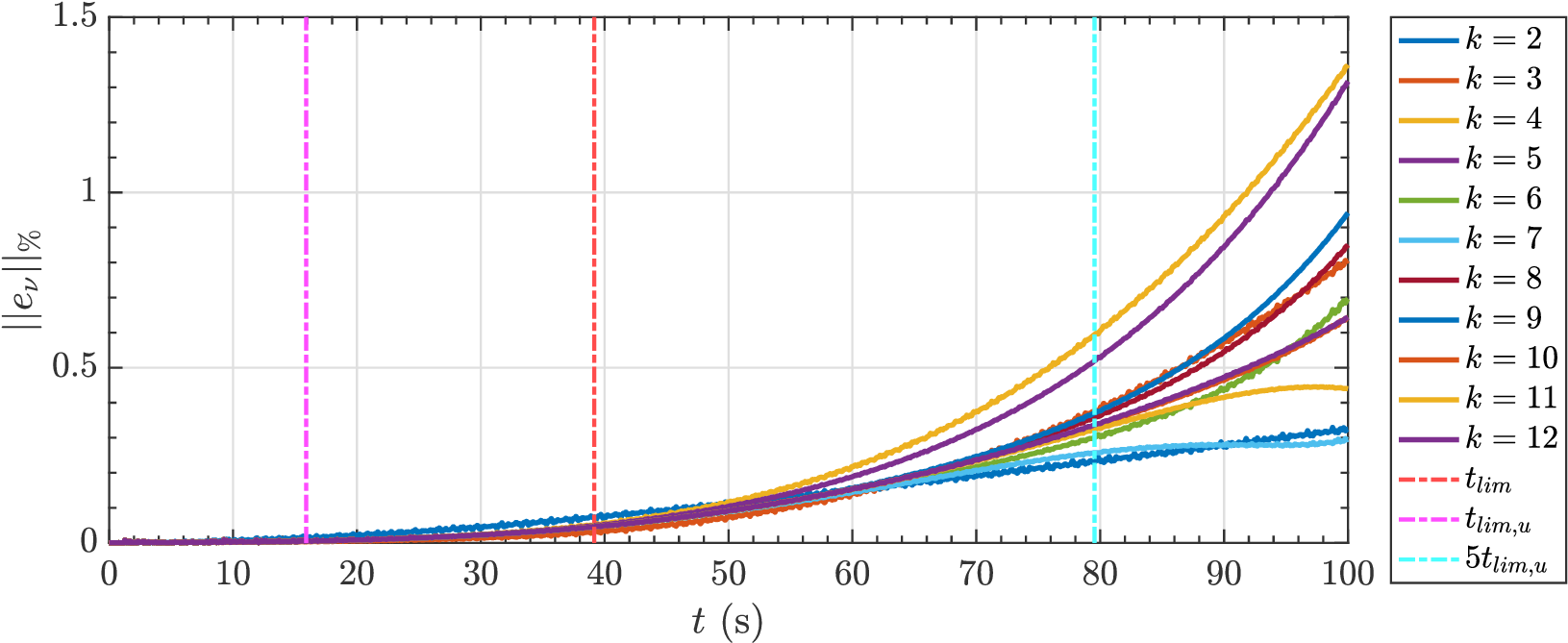}\hfill
\caption{\label{fig:13} Forced approximation error with $J=J_0$, $\nu(t_0) = 0.01$rad/s, and $\alpha = 10^{-3}$.}
\end{figure}

\begin{figure}[htb!]
\centering
\includegraphics[width=1\columnwidth]{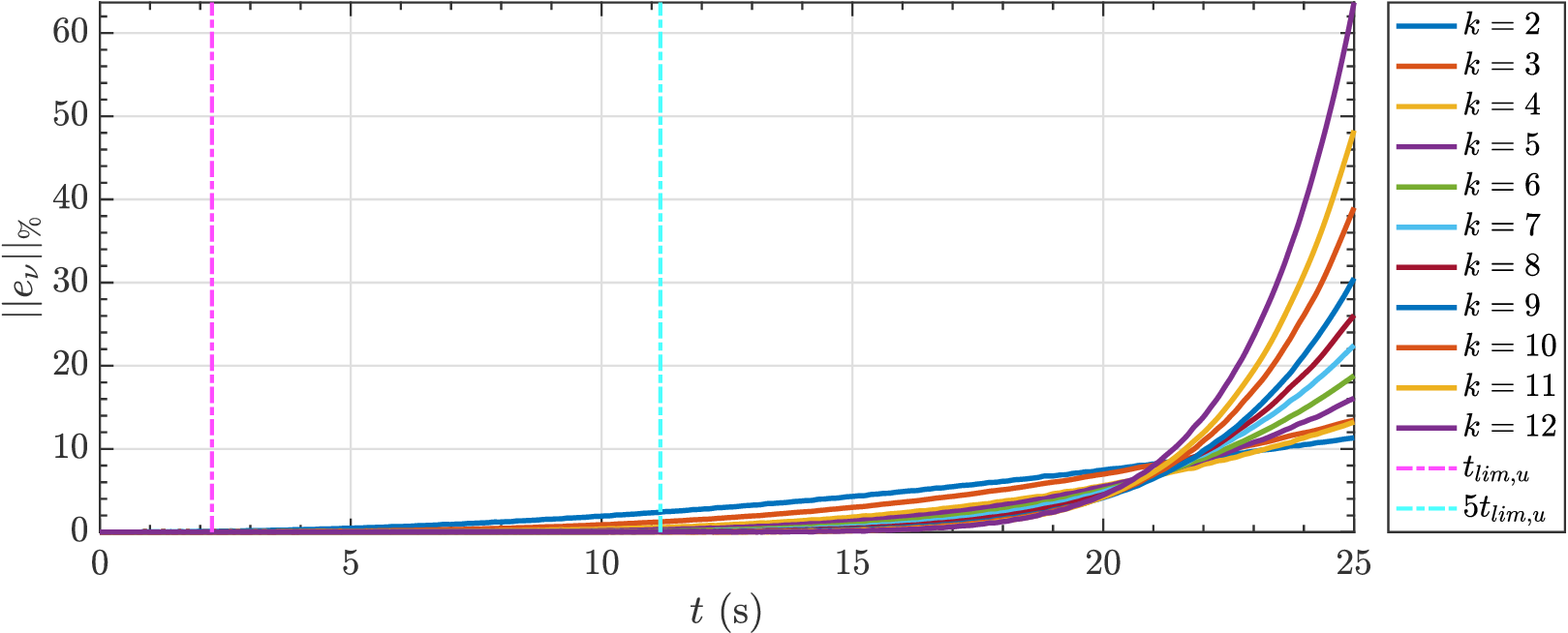}\hfill
\caption{\label{fig:14} Forced approximation error with $J=J_0$, $\nu(t_0) = 0.01$rad/s, and $\alpha = 10^{-2}$.}
\end{figure}

Figure \ref{fig:10}, \ref{fig:11}, \ref{fig:12}, \ref{fig:13}, and \ref{fig:14}, display the forced input response with $J = J_0$, $\nu(t_0)=0.001$ rad/s and $M = \alpha\left[\sin(2\pi t),\sin(2\pi(t-\pi/2)),\sin(2\pi t)\right]^{\top}$ with $\alpha=10^{-6},10^{-5},10^{-4},10^{-3},10^{-2},$ respectively. When computing $t_{lim,M}$, for simplicity, $\norm{\gamma(t)}$ is computed as the root mean square (RMS) of $\norm{J\nu_{nonlinear}}$, and $\norm{M_{\int}(t_0)}\approx\frac{\alpha}{2\pi}\sqrt{3}$. As expected from the results in section \ref{subsec:bounds}, with lower values of $\norm{M}$, such that $\frac{\norm{M}}{\norm{\gamma}}\ll\norm{J^{-1}}\norm{\gamma}$, the behavior resembles the one of the unforced system and the and the Koopman based model provides a good approximation of the nonlinear dynamics for $t<5t_{lim}$ (Fig. \ref{fig:10}). Contrariwise, when $\frac{\norm{M}}{\norm{\gamma}}\approx\norm{J^{-1}}\norm{\gamma}$, a good approximation is achieved for $t<5t_{lim,M}$, however the value of $t_{lim}$ is close to the value of $t_{lim,M}$ (Fig. \ref{fig:11}, \ref{fig:12}). Finally, when the $\frac{\norm{M}}{\norm{\gamma}}\gg\norm{J^{-1}}\norm{\gamma}$, a good approximation is achieved for $t<t_{lim,M}$.
\subsection{Position and Attitude Dynamics Numerical Validation}\label{subsec:fullvalid}
\begin{figure}[htb!]
\centering
\includegraphics[width=0.9\columnwidth]{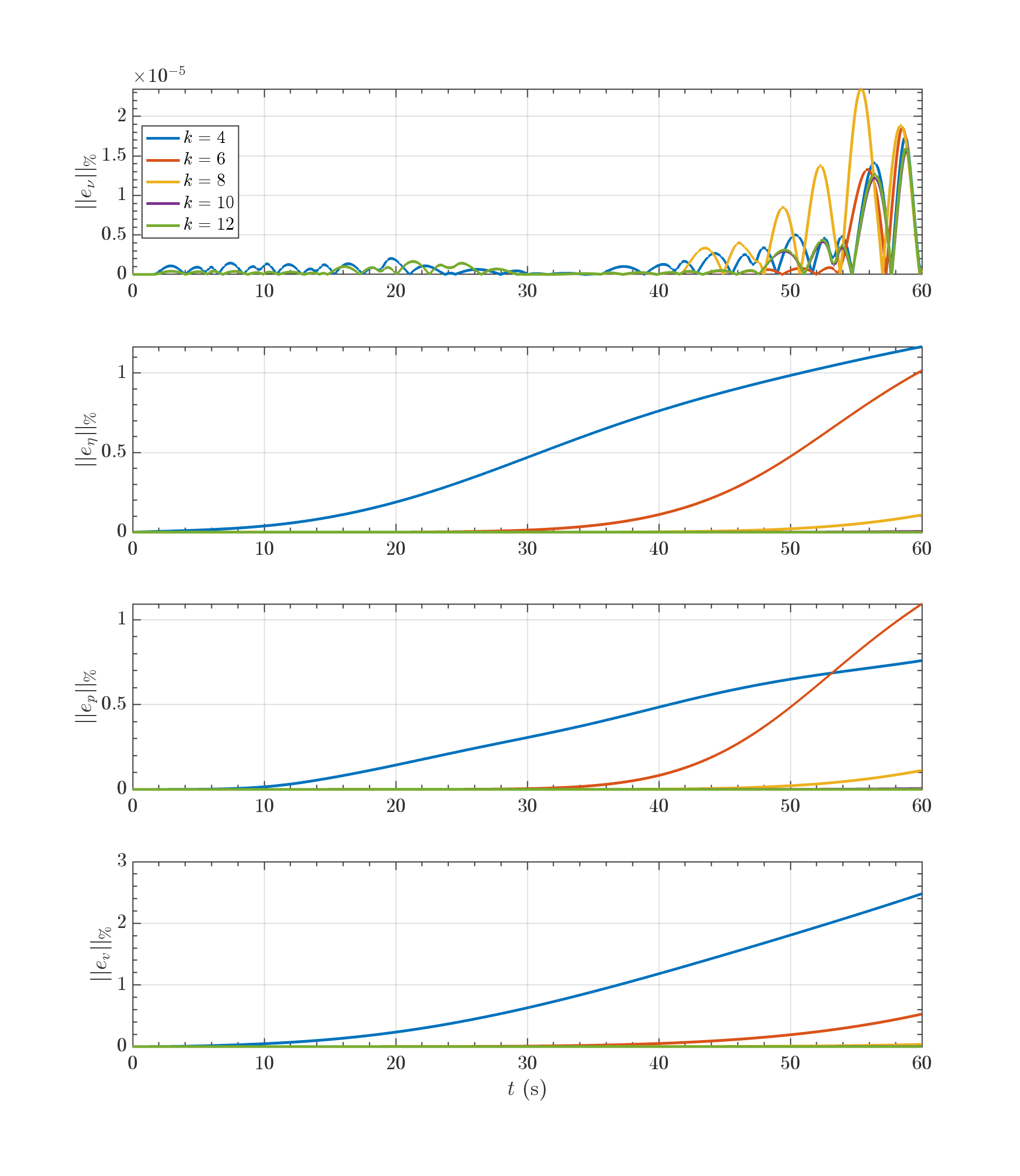}\hfill
\caption{\label{fig:16} Novel Koopman based position and attitude dynamics approximation error with respect to the number of observables with $N_{\nu}=N_{z}=k$, $J=I$.}
\end{figure}

Taking into consideration the limits found for the attitude model, in this subsection, the full position and attitude dynamics model is tested. First, in Fig. \ref{fig:16}, the position Koopman model based from \cite{martini2023koopman} is tested and compared to the lower order one introduced in \ref{subsec:LOKoop}. Both Model integrate the  Koopman based attitude dynamics tested above however the inertia tensor is set to $J = J_I=I_3$ so that the angular nonlinear dynamics reduces to linear one, only $k_{\nu}=1$ is used, and effectively rendering the test analog to the one in \cite{Chen,martini2023koopman,Zinage}. The model presents the expected increase of approximation of the nonlinear dynamics with the increase of the number of observables. Moreover, as shown in Table \ref{tab:compare}, the novel formulation presented in this work allows to use higher oreder observables while drastically reducing the system dimension and achieving an improved overall approximation with respect to \cite{martini2023koopman} and upon \cite{Zinage}.

\begin{table}[h]
\caption{Comparison with Prior Methods}\label{tab:compare}
\begin{center}
\begin{tabular}{cccc}
\hline
$N$      & $N = 39$         &\cite{martini2023koopman} $N=66$           &\cite{Zinage} $N=479$        \\ \hline
$\max\norm{p}_{\%}$                  & $2.048 \times 10^{-4}$ & $9.888 \times 10^{-4}$ & $4.923 \times 10^{-3}$ \\ 
$\max\norm{v}_{\%}$                   & $1.110 \times 10^{-4}$ & $3.341 \times 10^{-4}$ & $4.167 \times 10^{-3}$ \\ 
$\max\norm{\eta}_{\%}$  & $5.489 \times 10^{-4}$ & $5.850 \times 10^{-5}$ & $5.247 \times 10^{-4}$  \\ \hline
$\norm{\cdot}_{tot}$   & $5.963\times 10^{-4}$    & $1.045\times 10^{-3}$         &$6.471\times 10^{-3}$                  \\ \hline
\end{tabular}
\end{center}
\end{table}

\begin{figure}[htb!]
\centering
\includegraphics[width=0.9\columnwidth]{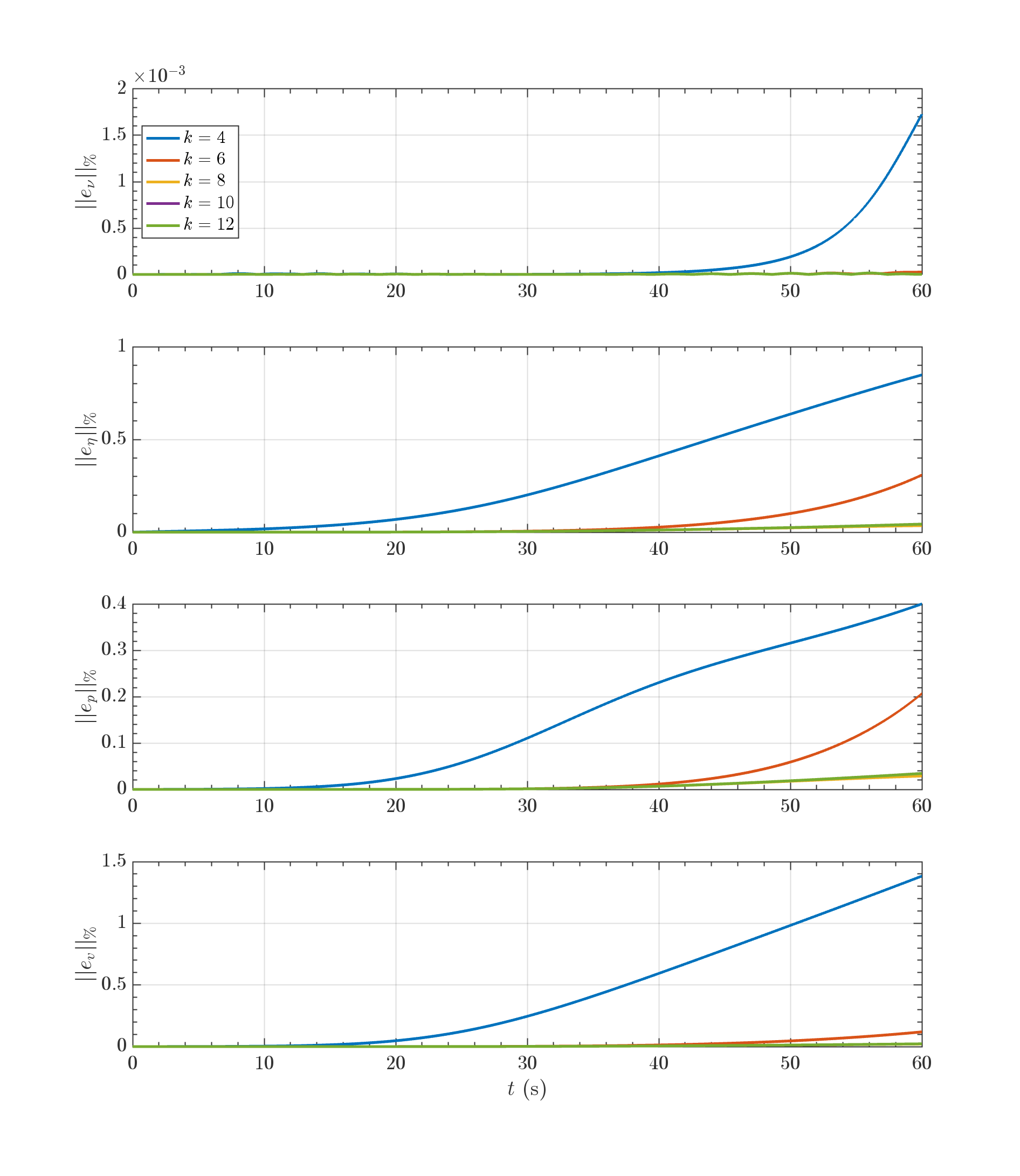}\hfill
\caption{\label{fig:17} Novel Koopman based position and attitude dynamics approximation error with respect to the number of observables with $N_{\nu}=N_{z}=k$, $J=J_0$, $\alpha = 10^{-5}$, $\beta_i = 1$, $\rho_i = 0.1$, and $\rho_t = 0$.}
\end{figure}

\begin{figure}[htb!]
\centering
\includegraphics[width=0.9\columnwidth]{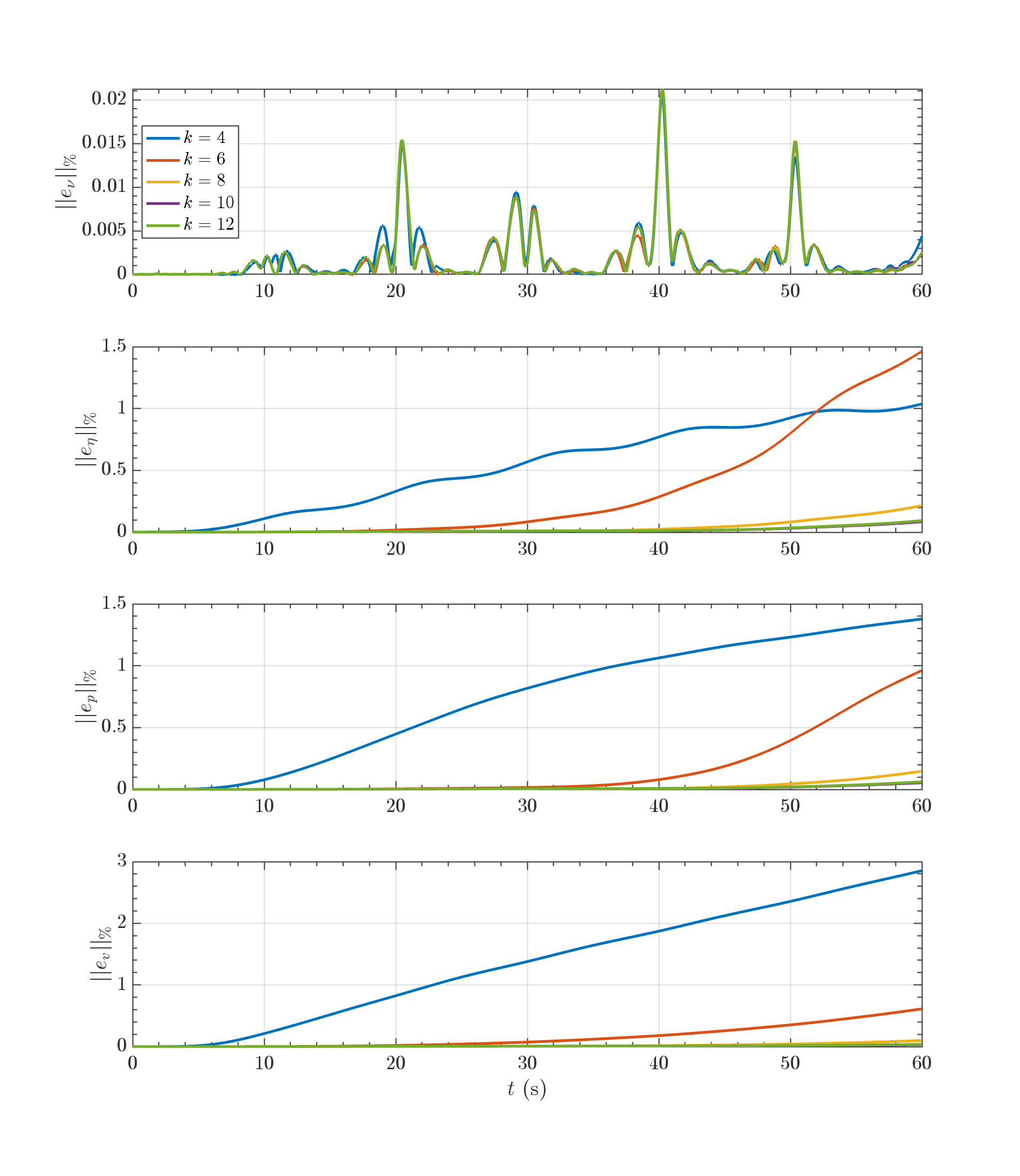}\hfill
\caption{\label{fig:18} Novel Koopman based position and attitude dynamics approximation error with respect to the number of observables with $N_{\nu}=N_{z}=k$, $J=J_0$, $\alpha = 10^{-4}$, $\beta_1 = 1$, $\beta_2 = 0.5$, $\beta_3 = 0.1$, $\rho_i = 2\pi10^{-1}$, and $\rho_t = \frac{\pi}{2}$.}
\end{figure}

\begin{figure}[htb!]
\centering
\includegraphics[width=0.9\columnwidth]{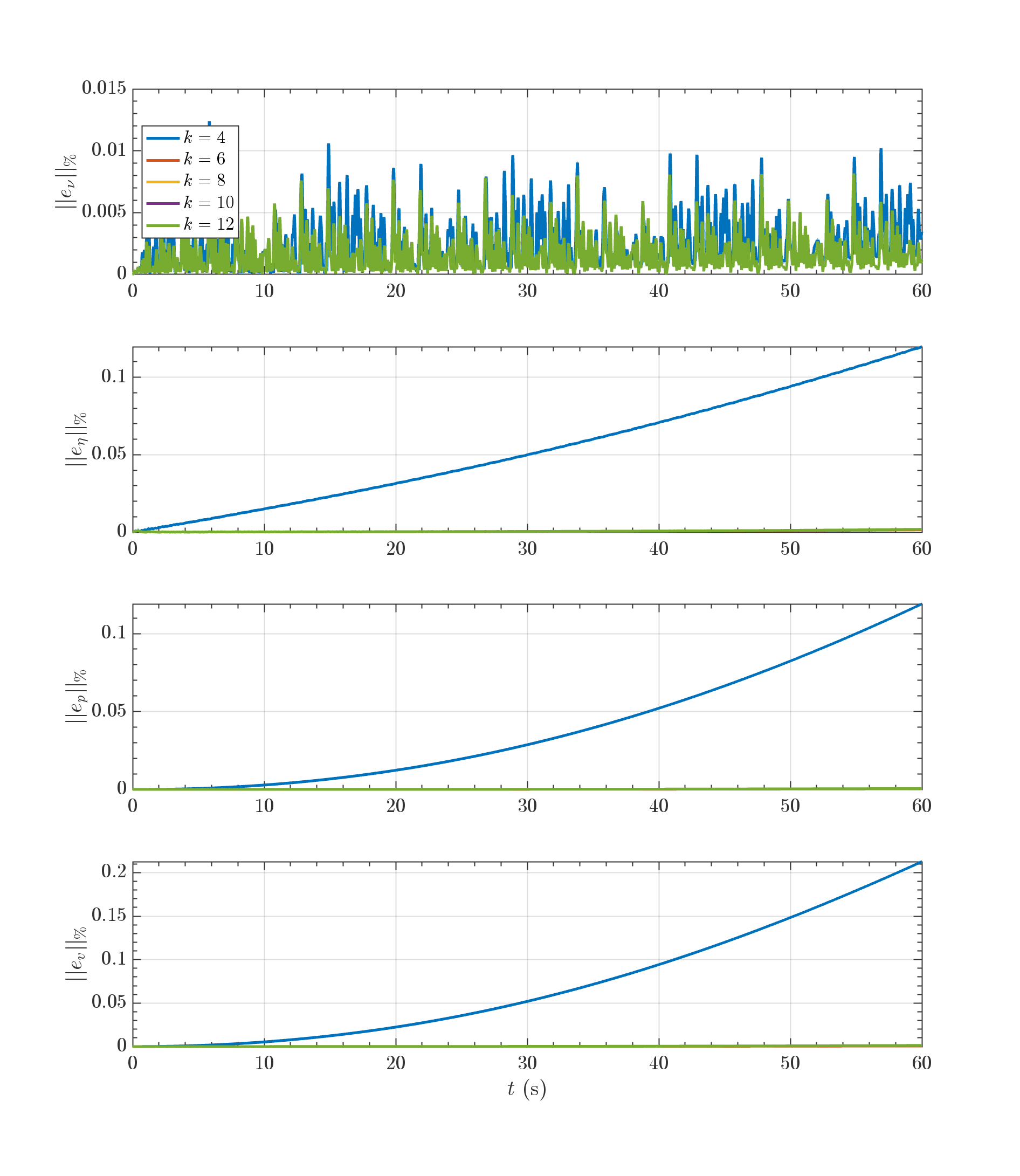}\hfill
\caption{\label{fig:19} Novel Koopman based position and attitude dynamics approximation error with respect to the number of observables with $N_{\nu}=N_{z}=k$, $J=J_0$, $\alpha = 10^{-4}$, $\beta_1 = 1$, $\beta_2 = 0.5$, $\beta_3 = 0.1$, $\rho_i = 2\pi$, and $\rho_t = \frac{\pi}{2}$.}
\end{figure}

\begin{figure}[htb!]
\centering
\includegraphics[width=0.9\columnwidth]{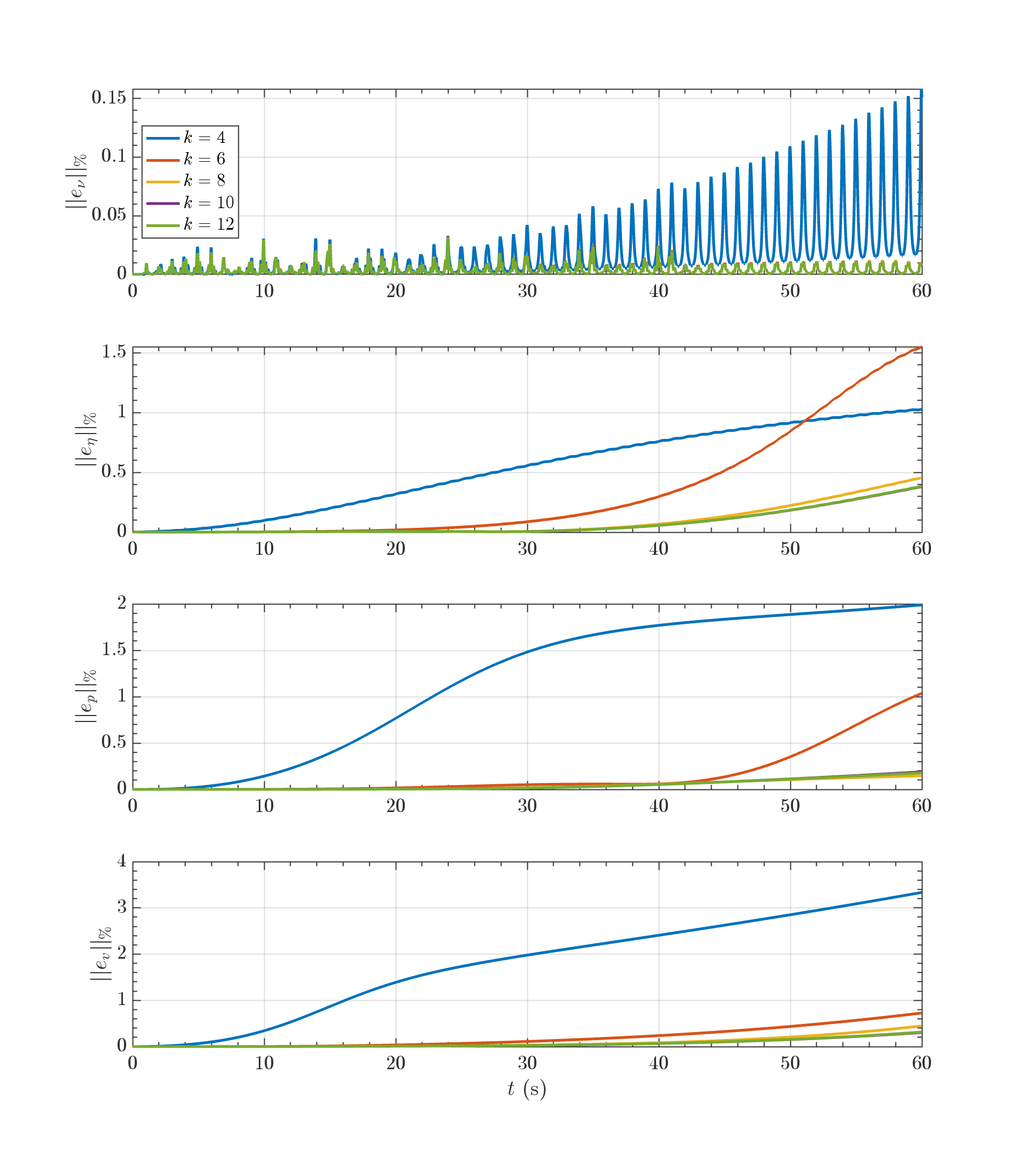}\hfill
\caption{\label{fig:20} Novel Koopman based position and attitude dynamics approximation error with respect to the number of observables with $N_{\nu}=N_{z}=k$, $J=J_0$, $\alpha = 10^{-3}$, $\beta_1 = 1$, $\beta_2 = 0.5$, $\beta_3 = 0.1$, $\rho_i = 2\pi$, and $\rho_t = \frac{\pi}{2}$.}
\end{figure}

Having established the improved behavior of the novel lower order Koopman based position and attitude dynamics, further tests are carried on with several values of inputs. Fig. \ref{fig:17}, \ref{fig:18}, \ref{fig:19}, \ref{fig:20} display the tests with $J = J_0$ with $M(t) = \alpha\left[\beta_1 \sin(\rho_1 t),\beta_2 \sin(\rho_2 (t-\rho_t),\beta_3 \sin(\rho_3 t)\right]^{\top}$, $F=\left[1,1,m*9.85\right]^{\top}$, and $\zeta= \left[M,F\right]^{\mathcal{T}}$. From the figures, it can be inferred that, for the used system parameters, a good approximation of the position and attitude dynamics is achieved with $k\geq8$. However, as previously seen for the attitude dynamics, the response is highly influenced from the magnitude and frequency of the input coupled with the initial conditions, and the time interval of a good model approximation drastically decrease when $\alpha \gtrapprox 10^{-2}$.
\subsection{Controllability Check}
From the analysis carried on in Section \ref{sec:3}, the matrices of \eqref{eqn:KoopAttPosJor} can be transformed in Jordan form, leading to six Jordan blocks associated to a zero eigenvalue, with same multiplicity. This decomposition can be interpreted as if each Jordan block is associated to the dynamics along each coordinate which, in the case of the Koopman based Position and attitude dynamics, can be expressed as the three components of the angular velocity $\nu_0$ and the three components of the coupled `rotated position' $z_0 = p_0$. Considering $J = J_0$, every case up to $N_{\nu} = N_{z} = 12$, it can be symbolically verified  that \eqref{eqn:ContrCheck} returns linearly independent rows, hence the system is fully controllable for $\nu\neq 0$.\\

\section{Quadrotor Koopman Formulation}\label{sec:6}
Considering now the simplified quadrotor model, the input force is composed only of the thrust in the body frame $z$-axis
\begin{equation}
    F = T\mathbf{e_3}
\end{equation}
while a the torque around all axes is produced by the thrust differential. Hence, a quadrotor has effectively four input to control the six degrees of freedom (DOF) of position and attitude. The resulting underactuation problem is generally tackled by dividing the feedback control loop in inner and outer control loop which, when deployed on hardware, run on different timestamps. The usual architecture employs the outerloop for position control, generating the desired thrust and reference attitude, while the inner loop computes the desired torques to guarantees attitude tracking. Considering now the formulation \eqref{complete_ss}, the recursive differential procedure carried out in section \ref{sec:Kpd} is such that the interdependence of the position and attitude dynamics is exposed in the control matrix, allowing, once the Koopman formulation is adapted to the quadrotor model, to control the observables dynamics and resulting in effective position control in the original state involving a single feedback control loops. 
To this end, in this section, the quadrotor Koopman model is presented, the approximation of the truncated lifted dynamics analyzed and, after the controllability check, a controller in the lifted dynamics is presented and tested in numerical simulations.

\subsection{Drone Model Validation}
\begin{figure}[htb!]
\centering
\includegraphics[width=0.9\columnwidth]{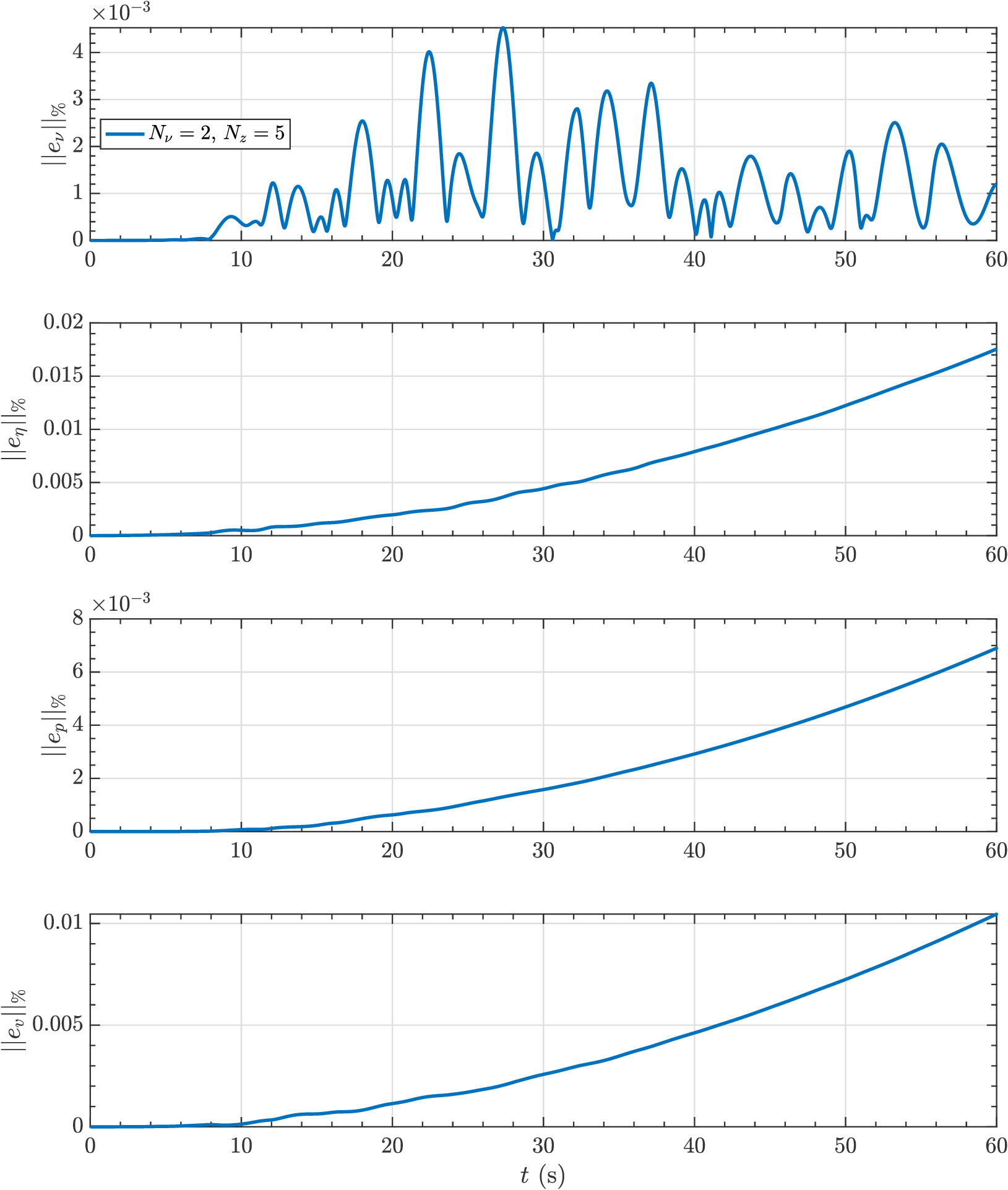}\hfill
\caption{\label{fig:21} Koopman based quadrotor model approximation error with number of observables $N_{\nu}= 2, N_{z}=5$, $J=J_Q$, $\alpha = 2\cdot10^{-6}$, $\beta_1 = 1$, $\beta_2 = 0.5$, $\beta_3 = 0.1$, $\rho_i = 2\pi10^{-1}$, and $\rho_t = \frac{\pi}{2}$.}
\end{figure}

The selected drone platform has the parameters listed in table $J_Q = diag (0.0131, 0.0131, 0.0234)$,. Given the drone parameters, it is expected a torque input of around $10^{-7} \sim 10^{-6}$ N/m and thrust of $10\sim 10^2$ N. Fig. \ref{fig:21} shows that Koopman model with a $N_{\nu} = 2$ and $N_z = 5$ gives a good approximation of the nonlinear dynamics. 

\subsection{Controller Design}
Along the same reasoning for which the complete model can be decomposed in six Jordan block associated to the six coordinates of $\nu_0$ and $z_0$, assuming that the angular velocity is available as measurement from the plant and, by restricting the control problem to the control of three spatial coordinates ($x,y,z$) and only one attitude coordinate (yaw angle $\psi$), it is plausible to reduce the system by employing only the third component of $\nu_k$ in addition to the full components of $z_k$. Moreover, considering the general symmetric structure of the quadrotor for which $J = diag(I_{x},I_y,I_z)$ where $Ix\approx Iy$, leads to $\nu_k(3)=0$ for $k>0$. Therefore, reduced model adopted for the controller design is 
\begin{align}
     x = (\nu_0(3),z_0,\{z_k\}^{5-1}_{k=1})\,\,\, \in\mathbb{R}^{N\times1}\,\,\, N = 16
\end{align}
which evolution is given by
\begin{align}
&\dot x = Ax + B(x)\zeta \label{eqn:complete_Koop_drone_ss}\\
&A = \left [
    \begin{array}{ccc}
[A_z] & 0\\
0 & 0
    \end{array}
    \right], \;B(x) = \left [
\begin{array}{cc}
\Xi_0 & -Z_0\\
\vdots & \vdots\\
\Xi_{4} & -Z_{4}\\
0_3 & 1/I_z
\end{array}
\right] \label{eqn:KoopDrone}
\end{align}
\subsubsection{Controllability Check}
Since the rows of 
\begin{align}
    \left [
\begin{array}{cc}
\Xi_{4} & -Z_{4}\\
0_3 & 1/I_z
\end{array}
\right]
\end{align}
are linearly independent, the resulting system is fully controllable. Compared to the original nonlinear system, this reduced system is fully controllable due to the coupling of position and attitude into a single observable which allows all the input to reach all the states in this new nonlinear coordinate frame. An important note is highlighting the fact that the there are multiple solution of Euler angles and position coordinates for a single value of $p_0$, however this can be solved by exploiting several higher order observables $p_k$ which will guarantee a single solution to be found.

For designing a control strategy, the quadrotor system of \eqref{eqn:complete_Koop_drone_ss} is rewritten as
\begin{eqnarray}\label{linear_dyn}
    \dot x &=& Ax + B^{\star}U^{\star}(x) \;,
\end{eqnarray}
where $U^{\star}(x) = B(x)\zeta$ and $B^{\star} = I_N$. Allowing to formulate the controller on a LTI system, while still guaranteeing controllability of the pair $(A,B^{\star})$. The system input is then computed by solving the least square optimization problem:
\begin{equation}\label{eq:LSOpt}
       \textbf{min:} (B(x)\zeta-B^{\star}U^{\star}(x))^{\top}(B(x)\zeta-B^{\star}U^{\star}(x)) 
\end{equation}
which solution is
\begin{eqnarray}
       \zeta &=& B^{\dagger}(x)B^{\star}U^{\star}(x) \;.
\end{eqnarray}
To guarantee the validity of such procedure it is crucial that the optimization error $B(x)\zeta-B^{\star}U^{\star}(x)$ remains sufficiently small during operation.\\
Tackling the trajectory tracking problem, a linear quadratic controller with integral action (LQI) \cite{young1972approach} is designed exploiting the Koopman based position and attitude dynamics of the quadrotor. As a first step, let's define the vector
\begin{align}
    x_e = &= (x_d-x)\\
    x_i &= \int^{t_f}_{t_0}x_e dt\\
    X &= [-x_e, x_i]^{\mathcal{T}}
\end{align}
the control input for the trajectory tracking controller is then obtained by minimizing the cost function 
 \begin{equation}
       J = {\frac{1}{2}} \int_{0}^{\infty} (X^{\top}QX + {U^{\star}}^{\top}RU^{\star}) \,dt \;,
\end{equation}
which solution is given by
\begin{equation}
           U^{\star} = -R^{-1}{B^{\star}}^{\top}P^{\star}X(t) = -K^{\star}X(t) + U^{\star}_d
\end{equation} 
Where $P \in \mathbb{R}^{N\times N}$ is the solution of the associated algebric Riccati equation, $R\in \mathbb{R}^{N\times N}$ and $Q \in \mathbb{R}^{N\times N}$ are weighting matrices to be tuned, and $U^{\star}_d$ is the desired input at steady state.
Finally the control input is computed as
\begin{equation}
    \zeta = B^{\dagger}(x)B^{\star}(-K^{\star}X(t) + U^{\star}_d).
\end{equation}
\subsection{Simulation}
The numerical simulations are performed in matlab simulink, the square trajectory to be tracked is computed by stitching of ninth order polynomial and then converted in Koopman observables coordinates. The simulation time is $70$ s with the quadrotr plant and controller running at a sampling rate of $1$ ms and $10$ ms respectively. The quadrotor starts at zero initial conditions with mass set to $1.2$ kg and $J=J_Q$.
\begin{figure}[htb!]
\centering
\includegraphics[width=0.9\columnwidth]{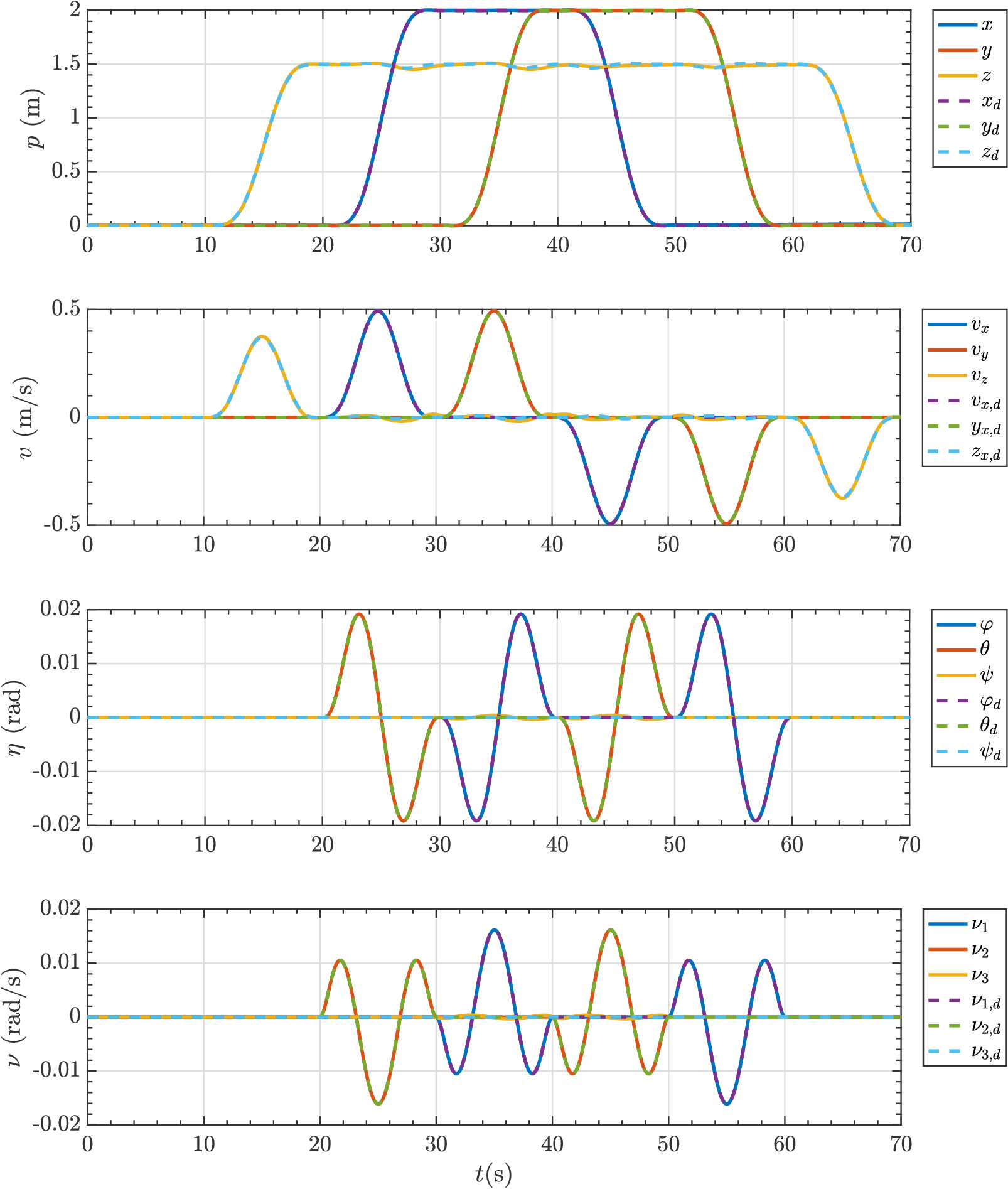}\hfill
\caption{\label{fig:22} Quadrotor actual and desired trajectory in time.}
\end{figure}
\begin{figure}[htb!]
\centering
\includegraphics[width=0.9\columnwidth]{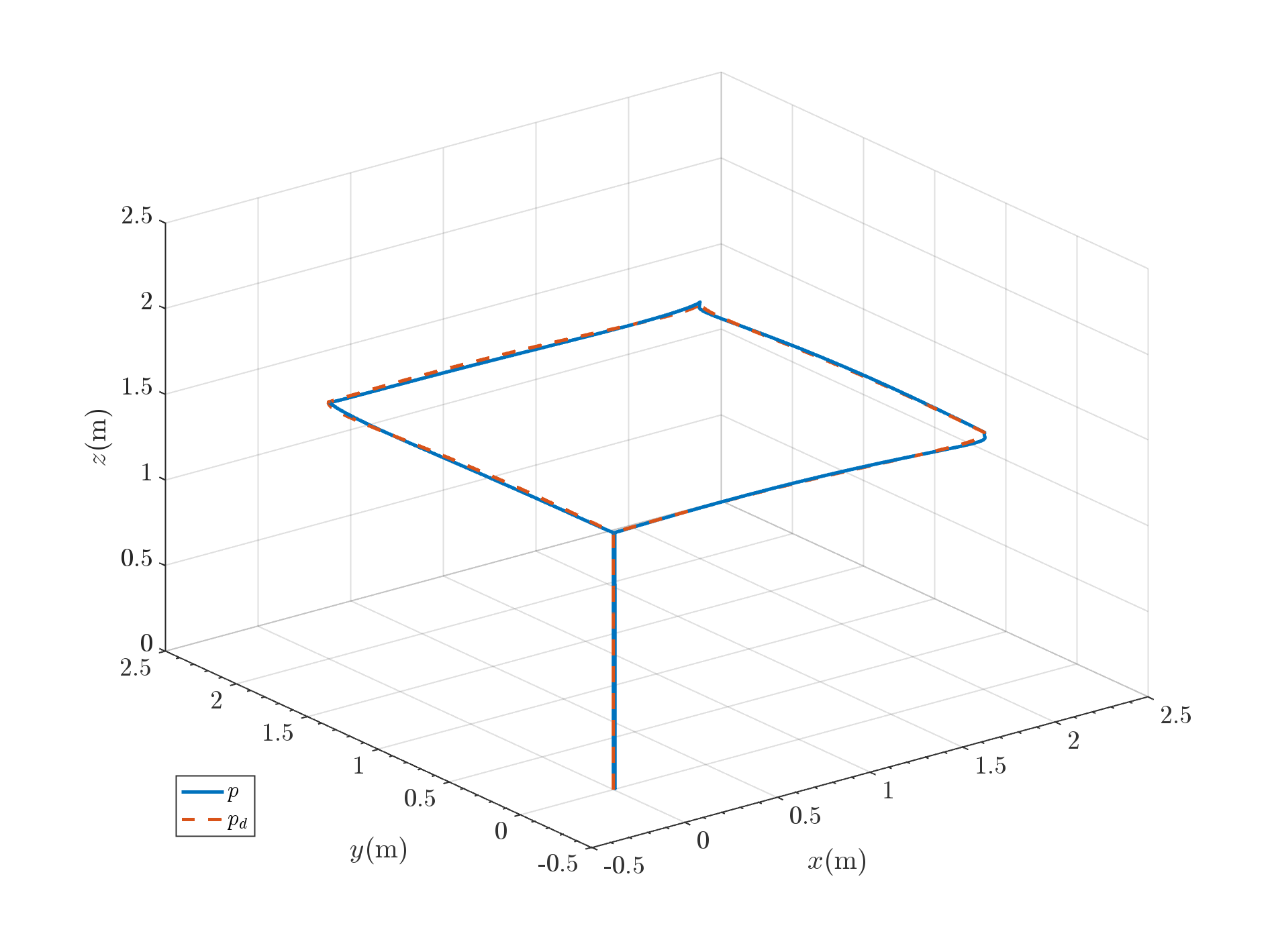}\hfill
\caption{\label{fig:23} Quadrotor actual and desired $x,y,z$ trajectory.}
\end{figure}
\begin{figure}[htb!]
\centering
\includegraphics[width=0.9\columnwidth]{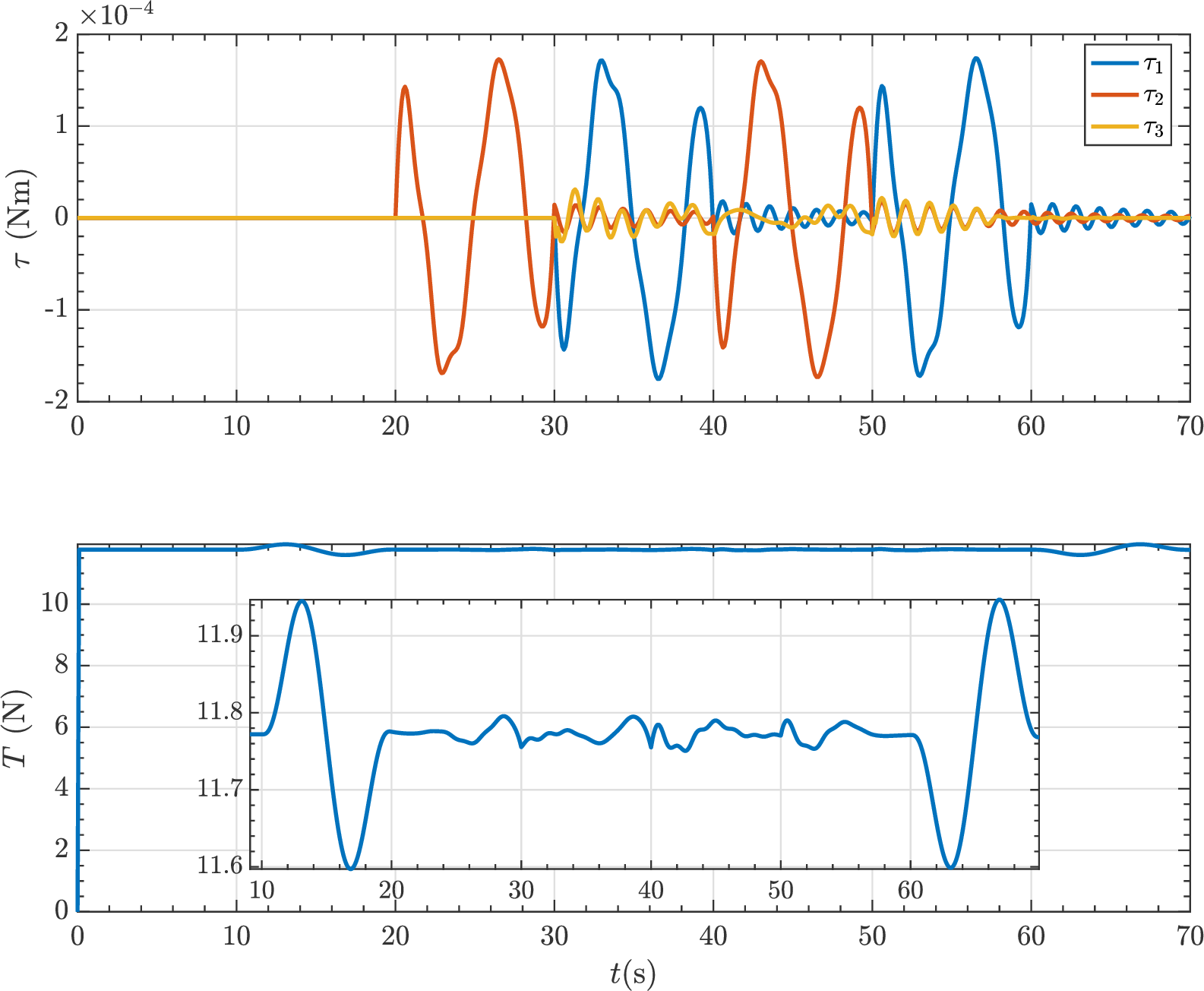}\hfill
\caption{\label{fig:24} Koopman based control action.}
\end{figure}
\begin{figure}[htb!]
\centering
\includegraphics[width=0.9\columnwidth]{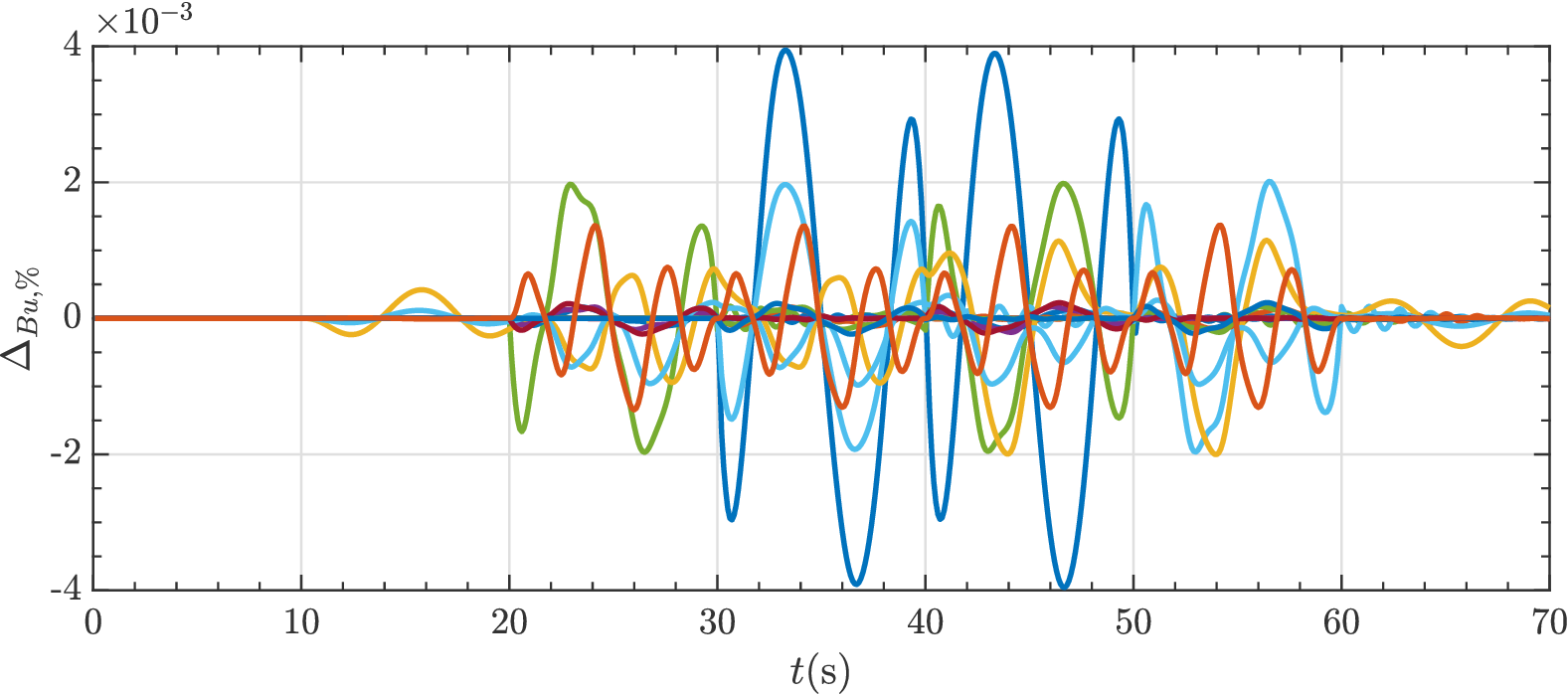}\hfill
\caption{\label{fig:25} Relative optimization error.}
\end{figure}

To achieve desirable results, the hand tuning process of the controller weighting matrices aims at minimizing the tracking error while maintaining a sufficiently low relative error of the least square optimization problem in \eqref{eq:LSOpt}, computed as 
\begin{equation}
    \Delta_{Bu,\%} = \frac{B(x)\zeta-B^{\star}U^{\star}(x)}{\norm{B(x)\zeta}}
\end{equation}

The simulation results in Fig. \ref{fig:22}, \ref{fig:23}, \ref{fig:24}, and \ref{fig:25}, show that the controller is able to successfully tracks the desired trajectory and the commanded actions remain bounded with sufficiently small optimization error for the entirety of the simulation.
\section{Conclusions}\label{sec:7}
A novel set of Koopman generalized eigenfunctions to embed the nonlinear position and attitude dynamics of a rigid body is presented. To the best of the author's knowledge, this is the first analytical computation of Koopman generalized eigenfunctions of attitude and position dynamics which does not require dynamic compensation of the angular velocity. Computing a truncated subset of the generalized Koopman eigenfunctions yields to a quasi linear dynamical system with constant state matrix in Jordan form and a state dependent control matrix. A controllability analysis provides an ad hoc controllability check to be performed on the truncated system and a boundary study returns a relaxed time interval for which the open loop numerical simulation of the derived model returns a good approximation of the original nonlinear dynamics. Compared to existing literature work, the presented novel formulation achieves a better approximation while maintaining a drastically more compact system dimension. The Koopman based model is then employed for the control system design of an underactuated quadrotor UAV. Results shows that the coupled nature of the Koopman generalized eigenfunctions allows for successful trajectory tracking using a single control loop, effectively solving the underactuation problem.

\end{document}